\documentclass[reprint,amsmath,amssymb,aps,prb,groupedaddress,nofootinbib,twocolumn,superscriptaddress]{revtex4-2}
\usepackage[defaultcolor=magenta]{changes}
\usepackage{graphicx}
\usepackage{amsthm,amssymb,amsmath,braket,mathdots}
\usepackage{bm}
\usepackage[pagebackref=false,pdfnewwindow=true]{hyperref} 
\usepackage{epstopdf,psfrag}
\usepackage{relsize,amsbsy}
\usepackage[export]{adjustbox}

\usepackage{tikz} 
\usetikzlibrary{arrows.meta}
\newcommand{\tstar}[5]{
\pgfmathsetmacro{\starangle}{360/#3}
\draw[#5] (#4:#1)
\foreach \x in {1,...,#3}
{--(#4+\x*\starangle-\starangle/2:#2) -- (#4+\x*\starangle:#1)}
-- cycle;}

\usepackage{graphicx,xcolor} 
\usepackage{tabularx, booktabs} 

\usepackage{units} 
\usepackage{dsfont} 

\DeclareMathOperator{\sgn}{sgn}
\DeclareMathOperator{\tr}{Tr}
\let\Re\relax 
\DeclareMathOperator{\Re}{\mathfrak{Re}}
\let\Im\relax
\DeclareMathOperator{\Im}{\mathfrak{Im}}

\newcommand{\ii}{{\mathrm{i}}} 
\newcommand{\e}{{\mathrm{e}}} 
\renewcommand{\v}[1]{\bm{#1}} 

\newcommand{\dd}{\partial}
\renewcommand{\d}{\mathrm{d}}
\newcommand{\1}{\mathds{1}} 
\newcommand{\trans}{\mathrm{T}}

\newcommand{\mac}{\mathcal}

\renewcommand{\eqref}[1]{Eq.~(\ref{#1})}

\definecolor{bananayellow}{rgb}{1.0, 0.88, 0.21}
\definecolor{straw}{rgb}{0.32, 0.28, 0.1}
\definecolor{orange}{rgb}{1.0, 0.53, 0.}

\def\cred{\color{black}} 

\begin{document}
\title{Theory of Difference Frequency Quantum Oscillations}
\author{V. Leeb}
\affiliation{Technical University of Munich, TUM School of Natural Sciences, Physics Department, 85748 Garching, Germany}
\affiliation{Munich Center for Quantum Science and Technology (MCQST), Schellingstr. 4, 80799 M{\"u}nchen, Germany}
\author{J. Knolle}
\affiliation{Technical University of Munich, TUM School of Natural Sciences, Physics Department, 85748 Garching, Germany}
\affiliation{Munich Center for Quantum Science and Technology (MCQST), Schellingstr. 4, 80799 M{\"u}nchen, Germany}
\affiliation{Blackett Laboratory, Imperial College London, London SW7 2AZ, United Kingdom}

\date{\today}

\begin{abstract}
Quantum oscillations (QO) describe the periodic variation of physical observables  as a function of inverse magnetic field in metals. The Onsager relation connects the basic QO frequencies with the extremal areas of closed Fermi surface pockets, and the theory of magnetic breakdown explains the observation of sums of QO frequencies at high magnetic fields.  Here we develop a quantitative theory of  {\it difference frequency} QOs in two- and three-dimensional metals with multiple Fermi pockets with parabolic or linearly dispersing excitations. We show that a non-linear interband coupling, e.g. in the form of interband impurity scattering, can give rise to otherwise forbidden QO frequencies which can persist to much higher temperatures compared to the basis frequencies. We discuss the experimental implications of our findings for various material candidates, for example multi-fold fermion systems, and the relation to magneto intersubband oscillations known for coupled two-dimensional electron gases.
\end{abstract}
	
\maketitle

\section{Introduction}
Quantum oscillation measurements have been a standard tool to determine electronic properties of metals since their discovery in bismuth by de Haas and van Alphen in 1930~\cite{deHaas1930}. QO were first reported in the magnetization, but soon afterwards magneto-transport measurements, known as the Shubnikov--de Haas effect, proved to be quantitatively similar~\cite{Shubnikov1930} as both originate from the discreteness of Landau levels of electrons in a magnetic field~\cite{landau1930diamagnetismus}. Onsager later realized that the oscillation frequency of the magnetization or the conductivity as a function of the inverse magnetic field is directly related to the metal's Fermi surface~\cite{Onsager1952}, before Lifshitz and Kosevich (LK) completed the canonical theory of QOs by connecting the temperature dependence of the amplitude of QOs to the effective mass of the electrons~\cite{Lifshitz1956}. Hence, QO measurements can detect the size of even tiny Fermi pockets, the effective mass of electrons and the scattering rate via the Dingle temperature~\cite{Shoenberg1984}. 

Deviations to the standard theory of QO are rare and exotic. For example, the observation of anomalous QOs in bulk insulators~\cite{tan2015unconventional,hartstein2018fermi,liu2018fermi,xiang2018quantum,hartstein2020intrinsic} and heterostructures~\cite{xiao2019anomalous,han2019anomalous,wang2021landau,leeb2021anomalous} is believed to be a result of strong electron correlations and  has recently led to a flurry of new theoretical proposals beyond the standard LK theory~\cite{knolle2017anomalous,shen2018quantum,erten2016kondo,zhang2016quantum,sodemann2018quantum,lee2021quantum,he2021quantum}. In contrast, it is well established and long understood that QO frequencies beyond the simple Onsager rule can appear in strong magnetic fields~\cite{Cohen1961,Blount1962} where the basic semiclassical description of electrons, simply traveling along the edge of the cross sectional area of the Fermi surface, breaks down and tunneling between distinct Fermi pockets becomes important~\cite{slutskin1968dynamics}. In this magnetic breakdown theory a whole zoo of combinations of frequencies can arise, depending strongly on the gaps between different semiclassical orbits and Berry phase effects~\cite{alexandradinata2018semiclassical}. 

\begin{figure}
	\includegraphics[scale=1]{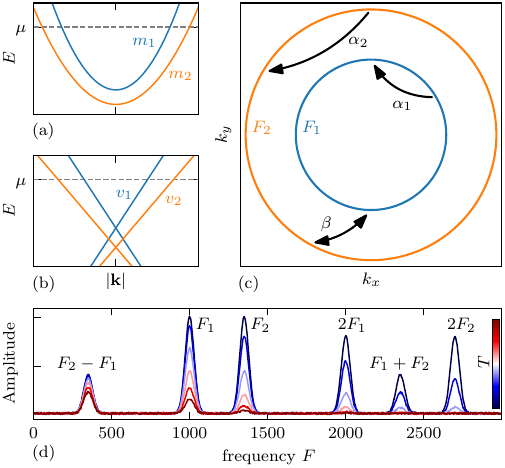}
	\caption{ 
	We consider generic two band models in 2D and 3D with either parabolic bands with effective masses $m_1$ and $m_2$,  panel (a), or relativistic bands with Fermi velocities $v_1$ and $v_2$, panel (b). Panel (c) shows the extremal cross section of the two resulting Fermi surface pockets (in 3D for fixed $k_z$). The two Fermi surfaces give rise to two distinct basic QO frequencies $F_1$ and $F_2$. A new difference frequency with $F_1-F_2$ appears via interband impurity scattering of strength  $\beta$ (the intraband scattering $\alpha_\lambda$ leads to a standard Dingle suppression). Panel (d) shows schematically the expected Fourier spectrum of the oscillating part of the conductivity for our effective two band systems. Remarkably, the difference frequency $F_1-F_2$ is stable for increasing temperature.}
	\label{fig1}
\end{figure}

A new QO frequency appears, for example, in \mbox{type-II} Weyl semimetals associated with the difference of two electron- and hole-type Fermi surface areas~\cite{OBrien2016,alexandradinata2017geometric,van2018electron}, which can be understood via magnetic breakdown in the form of Klein-tunneling between the counter-propagating semiclassical orbits. However, not all possible combinations of frequencies appear within the semiclassical breakdown theory, e.g. a {\it difference frequency} originating from two parallel electron-like (or two hole-like) pockets, see Fig.~\ref{fig1}~(c), seems impossible.     

In this work we study the effect of non-linear interband coupling on the oscillating part of the conductivity in multi-band metals. We show in detail how interband impurity scattering can lead to sum and difference frequencies in the Shubnikov--de Haas effect. Unlike  magnetic breakdown, the appearance of these frequencies is not induced by the magnetic field but triggered by self-energy effects originating from the non-linear coupling of distinct Landau quantized pockets.  Remarkably, the emerging difference frequency is often only weakly temperature damped such that it can persist to much higher temperatures then its semiclassical basis frequencies.

In the context of two dimensional (2D) electron gases (2DEG) a related phenomenon, dubbed {\it magneto intersubband oscillations}, has been studied previously~\cite{Polyanovsky1988,Raikh1994,Averkiev2001}. The unusual temperature stability is in accordance with experimental observations on 2DEGs~\cite{Coleridge1990,Leadley1992, Goran2009} and has also been predicted to appear in quasi-2D, layered metals~\cite{Polyanosky1993, Grigoriev2003, Thomas2008,Mogilyuk2018}. Here, we generalize the theory of QO {\it difference frequencies} to generic parabolic and linearly dispersing band structures and establish that they even appear for isotropic 3D systems, which is of experimental relevance to a number of materials classes, e.g. multifold fermion systems~\cite{manes2012existence,Bernevig2012,bradlyn2016beyond,tang2017multiple} as very recently observed in CoSi~\cite{Huber2023}.

\section{Summary of Results}
\label{sec:summary}
\subsection{Summary and Review}
We consider generic two band Hamiltonians in 2D and 3D, the exemplary band structures are shown in Fig.~\ref{fig1}~(a) and (b). The common feature of our models is the existence of two Fermi surfaces, see Fig.~\ref{fig1}~panel~(c). Note that, we do not expect any magnetic breakdown between the two pockets because the velocity of the semiclassical orbits has only parallel components. 
The key ingredient for the appearance of a difference frequency is a non-linear coupling of the bands which we study in terms of generic impurities. In addition to the standard intraband scattering channels $\alpha$, the interband  $\beta$-channel allows electrons to scatter between the two distinct Fermi surfaces, see Fig.~\ref{fig1}~(c). We concentrate on short-ranged impurities, which permit an analytical calculation of the conductivity and capture the main contribution for dominant $s$-wave scattering. 

The emergence of sum and difference frequencies can be seen by considering the generic formula of the conductivity
\begin{equation}
\sigma = \int_{-\infty}^\infty \d \varepsilon [-n_F'(\varepsilon)] \hat \sigma(\mu+\varepsilon),
\label{eq:convolution}
\end{equation}
which can be expressed as a convolution of the derivative of the Fermi distribution function $-n_F'(\varepsilon)= \frac{1}{4 T \cosh^2\left(\nicefrac{\varepsilon}{2T}\right)}$ (with chemical potential $\mu$ and temperature $T$) and the conductivity kernel $\hat \sigma(E)$. The kernel includes a sum of various contributions from the two bands and  different harmonics. Concentrating on the oscillating components only, one can write each as a product of a non-oscillating term $g$ and an oscillating function~\cite{Shoenberg1984}  as
\begin{equation}
\hat\sigma(E) = \cos(f(E)) g(E).
\label{eq:condKernel_genericForm}
\end{equation}
In this expression $f(\mu) = 2 \pi F/B + \phi$ includes the dependence on the cross sectional area of a Fermi pocket $F(\mu)$ and a phase $\phi$. 

First, let us recall how to obtain the canonical LK result for QOs via \eqref{eq:convolution} and~(\ref{eq:condKernel_genericForm}). Using the fact that the chemical potential $\mu$ is large compared to the temperature one expands the kernel in a region $k_B T$ around $E=\mu$. The resulting integral can be evaluated analytically, see Appendix A,  and one finds the standard behavior
\begin{equation}
\sigma = \cos(f(\mu)) g(\mu) R_\text{LK}(\pi f'(\mu, T)
\label{eq:cond_f}
\end{equation}
with the famous LK temperature dependence
\begin{equation}
R_\text{LK}(\chi) = \frac{\chi}{\sinh \chi}.
\label{eq:LK1}
\end{equation}
We note that higher orders in the expansion $k_B T/\mu$ can lead to non-LK behavior as we discuss in Appendix \ref{app:temperature convolution}.

Next, let us discuss the apperance of a difference frequency via impurity scattering. Intraband contributions can lead to oscillations in the normally assumed to be non-oscillating prefactor $g$, as well as to oscillations of the chemical potential \cite{Peres2006,Grigoriev2003}. However, the quantities always oscillate with the same frequency $F$ as the basic QO, i.e. their oscillations depend on the cross sectional area of the Fermi surface associated with the same band. Therefore, these perturbative effects can only change the non-universal part of the amplitude of the QOs.

The key observation is that due to the interband scattering channel $\beta$ the quantities $g$ and $\mu$ can also oscillate with the frequency associated with the {\it other} Fermi pocket! Thus, {\cred the basic oscillations of the conductivity in \eqref{eq:condKernel_genericForm} are modulated  by the oscillations of $g$.} In the second harmonic this leads to two new frequencies: the sum and the difference of the two basis frequencies
\begin{equation}
\sigma_{\pm} \propto \cos\left(2\pi \frac{F_1 \pm F_2}{B}\right) R_{D1} R_{D2} R_\text{LK}\left(2\pi^2 \frac{m_1 \pm m_2}{e B} T\right), 
\label{eq:cond_sum/diff}
\end{equation}
which is the main result of our work. 
Note, the sum and difference frequency resemble the standard LK form with the generic damping factor $R_\text{LK}\left(2\pi^2 \left(\frac{\dd F_1}{\dd \mu}-\frac{\dd F_2}{\dd \mu}\right) \frac{T}{B}\right)$, and both are a second order effect in the Dingle factor $R_{D}$, which describes damping from impurity scattering as discussed below. Hence, their Dingle temperature is in either case a sum of the two basis Dingle temperatures, weighted with their effective masses. Strikingly, the difference frequency can persist to much higher temperatures then the basis frequencies. Following \eqref{eq:cond_sum/diff} the sum and difference frequency decay for parabolic bands with the sum and the difference of the effective masses of the two bands. If the effective masses around the Fermi energy are  equal, e.g. the difference of the cross sectional areas of the Fermi surfaces does not change as a function of  the chemical potential, the difference frequency acquires no temperature smearing at all! Note that the temperature dependence of sum and difference frequency inverts for coupled electron/hole pockets.

For relativistic dispersions with Dirac/Weyl-like excitations similar expressions to \eqref{eq:cond_sum/diff} are obtained. For these linearly dispersive bands with Fermi velocity $v_\lambda$ the difference frequency remains however slightly temperature dependent even for equal Fermi velocities. The reason for this is the quadratic dependence of the extremal cross section of the Fermi surface on the chemical potential.

We note that beyond second order in the Dingle factor higher combination frequencies can appear. Any integer combination $k_1 F_1 + k_2 F_2$ with $k_1, k_2 \in \mathbb{Z}$ is allowed and comes with the Dingle factors of $R_{D1}^{|k_1|}R_{D2}^{|k_2|}$. Strikingly, all combination with negative $k_1,k_2$, including the difference frequency, are absent to leading order in the density of states and therefore also in the de Haas--van Alphen effect.

\subsection{Relation to earlier work}
The insight that interband scattering can lead to a sizable temperature stable difference frequency in the conductivity is well known for 2DEGs \cite{Polyanovsky1988,Raikh1994,Averkiev2001}. Furthermore, similar effects are known to appear in quasi-2D layered parabolic metals \cite{Polyanosky1993,Grigoriev2003,Thomas2008,Mogilyuk2018}, where impurities couple the different layers. Systems where nearly equal effective masses have lead to the observation of a temperature stable difference frequency include GaAs heterostructures \cite{Leadley1989,Coleridge1990,Leadley1992, Sander1996,Goran2009,Minkov2020}, metals with bilayer crystal structure \cite{Grigoriev2017,Grigoriev2016} and organic metals \cite{Kartsovnik2002}. Here, we first extend the 2D theory to Dirac systems with linear dispersions. The second and main new finding of our work is to establish that difference frequency oscillations may also appear for isotropic 3D systems with generic dispersions. Our work provides a framework for the {\it theory of difference frequency QOs} being applicable to generic band structures in any dimension. We also provide qualitative arguments for the behavior of higher harmonic frequencies and their unusual temperature dependencies.

\subsection{ Outline}
The remainder of our work is organized as follows: In section~III we first rederive the oscillating part of the conductivity for parabolic dispersions in 2D and obtain similar results as Ref.~\cite{Raikh1994}. We discuss in detail the interband scattering contribution to the self-energy, before we generalize the results to 3D. In section~IV we proceed with analogous calculations for relativistic fermions, e.g. effective descriptions of Dirac and Weyl/multifold fermion materials. In section~V we show that the difference frequency is to leading order absent in the de Haas--van Alphen effect. Finally, in section~VI we conclude with a discussion of our results and present exemplary  3D and/or Dirac materials in which we expect a temperature stable difference frequency to be observable.

\section{Parabolic dispersions}
\subsection{Two dimensions}
\label{sec:3A}
QOs in the conductivity can be seen in nearly every known metal disregarding its specific features like interactions or spin--orbit coupling. Hence, our models should be seen as effective descriptions of the excitations around the Fermi energy which emerge after incorporating all microscopic details. For simplicity we start by considering a generic two band Hamiltonian
\begin{equation}
H = \sum_{\v{k},\lambda} \epsilon_\lambda (\v{k}) c_{\v{k},\lambda}^\dag c_{\v{k},\lambda} + \sum_{\v{r}} U(\v{r}) \v{c}^\dag_{\v{r}} \Lambda \v{c}_{\v{r}}
\label{eq:2D_hamiltonian}
\end{equation}
in 2D. The quadratically dispersive bands $\lambda$ with dispersion $\epsilon_\lambda(\v{k}) = \frac{\v{k}^2}{2m_\lambda}-W_\lambda$ have different effective masses $m_\lambda$ and are shifted with respect to each other by $W_1-W_2$, see Fig.~\ref{fig1}~(a). They can in principle also be shifted in momentum space with respect to each other, modeling different electron or hole pockets. The electrons $\v{c}_{\v{r}} = (c_{\v{r},1},c_{\v{r},2})^\trans$ can scatter on impurities located at positions $\v{r}_i$. The impurities are distributed randomly and uniformly such that the systems remains on average translationally invariant which we model by the short-ranged potential 
$U(\v{r}) = U_0 \sum_{\v{r}_i} \delta \left(\v{r}-\v{r}_i \right)$. The key ingredient is the scattering vertex $\Lambda$ which has intraband channels $\alpha_\lambda$ and an interband channel $\beta$
\begin{equation}
\Lambda = 
\begin{pmatrix}
\sqrt{\alpha_1} & \sqrt{\beta} \\
\sqrt{\beta} & \sqrt{\alpha_2} \\
\end{pmatrix},
\label{eq:2D:scattering_vertex}
\end{equation}
and allows electrons to scatter between the distinct Fermi surfaces. The dimensionless numbers $\sqrt{\alpha_\lambda}$ and $\sqrt{\beta}$ quantify the effective scattering rates and we have absorbed the complex phase of $\sqrt{\beta}$ in the definitions of $c$ and $c^\dag$. The following calculation follows similar steps as in Ref.~\cite{Raikh1994}.

In order to study QOs, we introduce a quantizing magnetic field $\v{B} = B \hat e_z$, perpendicular to the 2D system. The vector potential is chosen in the Landau gauge $\v{A}=(-B y,0,0)^\trans$. Peierls substitution leads to LLs for each band of the form $\epsilon_\lambda(l) = \omega_{c\lambda} \left(l+\nicefrac{1}{2}\right)-W_\lambda$ where the cyclotron frequency is $\omega_{c\lambda} = \frac{e B}{m_\lambda}$. The field operators now carry the following quantum numbers: band index $\lambda$, LL index $l$ and the trivial momentum $k_x$. The wavefunctions are the usual ones of a shifted harmonic oscillator at $y_0 = \nicefrac{k_x}{eB}$.

\subsubsection{Conductivity}
Our objective is to compute QOs in the conductivity and to proceed analytically we concentrate on the transversal component  $\sigma_{xx}$. Following the Kubo formula \cite{Bastin1971} the conductivity kernel appearing in \eqref{eq:convolution} is given by
\begin{equation}
\hat\sigma_{xx}(E) = \frac{e^2}{\pi L_x L_y} \mathrm{Tr}_{l,k_x,\lambda}[v_x \Im G(E) v_x \Im G(E)]
\label{eq:general_Kubo}
\end{equation}
where $G(E)$ is the retarded, impurity averaged Green's function $G_{\lambda,l}(E) =p (E-\epsilon_\lambda(l)-\Sigma_\lambda(E))^{-1}$ and $v_x$ is the velocity operator. For short-range impurity scattering, the self-energy $\Sigma_\lambda$ does not depend on any of the electron quantum numbers except the band index $\lambda$, see section~\ref{sec:2D:self-energy}. Furthermore, we assume in the notation for $G$ that the self-energy remains diagonal in the band index $\lambda$ which is an approximation discussed below in section~\ref{sec:2D:self-energy}. 

In order to reduce the complexity of the notation in this manuscript, we will use the dimensionless energy $\xi_\lambda = \frac{E+W_\lambda}{\omega_{c\lambda}}$ which will always appear together with the real part of the self-energy and define $\xi_\lambda^\star = \xi_\lambda-\Re \Sigma_\lambda /\omega_{c\lambda}$. Note, because the Fermi distribution function $n_F'(\xi)$ is strongly peaked in a region $k_B T$ around $\mu$, and as $\nicefrac{\mu}{\omega_{c \lambda}} \gg 1$, we may take $\xi_\lambda^\star \rightarrow \infty$ for all integration boundaries. Furthermore, we denote the imaginary part of the dimensionless self-energy by $\Gamma_\lambda = -\nicefrac{\Im\Sigma_\lambda}{\omega_{c \lambda}}$ and introduce $\bar \lambda$ which takes the value 2 (1) if $\lambda$ is 1 (2). 

Inside a magnetic field the velocity operator is quantized as $v_x = \frac{\sqrt{e B}}{\sqrt{2} m_\lambda}\left(a^\dag + a\right)$, where $a^\dag$ and $a$ are the ladder operators of the shifted harmonic oscillator. After evaluation of the trace the conductivity kernel is
\begin{equation}
\hat\sigma_{xx}(E) = \sigma_0 \frac{N_\Phi}{2 L_x L_y} \sum_{\lambda,l=1} l \Im G_{\lambda,l}(E) \Im G_{\lambda,l-1}(E),
\label{eq:2D:cond_sum over l}
\end{equation}
with $\sigma_0 = \nicefrac{2 e^2}{\pi}$. The sum over Landau levels can be transformed into a sum over harmonics using the standard Poisson summation formula
\begin{equation}
\sum_{l=0}^\infty f(l) = \sum_{k=-\infty}^\infty \int_0^\infty \d x \e^{2\pi \ii k x} f(x).
\label{eq:poisson_sum}
\end{equation}
The resulting integral can be solved exactly by extending the lower boundary to $-\xi_\lambda^\star \rightarrow -\infty$ and then performing complex contour integration. The final conductivity kernel takes the form
\begin{align}
\hat \sigma_{xx} (E) =& \sigma_0 \sum_\lambda \frac{\xi_\lambda^\star |\Gamma_\lambda(\xi)|}{1+4\Gamma_\lambda(\xi)^2} \nonumber\\
&\times\left(1+ 2\sum_{k=1}^\infty (-1)^k \cos(2\pi k \xi_\lambda^\star) R_\lambda(\xi)^k\right).
\label{eq:2D:cond_result}
\end{align}
The conductivity can then be expanded as a power series in the damping factor
\begin{equation}
R_\lambda(\xi) = \exp\left(-2\pi |\Gamma_\lambda(\xi)| \right).
\label{eq:DampinfFac}
\end{equation}

The standard canonical QOs can now be recovered by setting $\Im \Sigma_\lambda$ to a constant, the empirical Dingle temperature $T_{D,\lambda}$, such that $R_\lambda(\xi)$ becomes the well known Dingle damping factor~\cite{Shoenberg1984}
\begin{equation}
R_{D,\lambda} = \exp\left(-2\pi^2 \frac{T_{D,\lambda}}{\omega_{c\lambda}} \right).
\label{eq:DingleDamping}
\end{equation}
One would then follow our discussion in section~\ref{sec:summary} (\eqref{eq:condKernel_genericForm} to \eqref{eq:cond_f}) to evaluate the convolution with the Fermi distribution function \eqref{eq:convolution} to obtain QOs of the well known LK form, which are also in accordance with the semiclassical Onsager relation. In the next section we will go beyond the simple assumption that the self-energy is a mere constant, but show that it can acquire oscillations with two basis frequencies due to interband impurity scattering.

%

\subsubsection{Self-energy}
\label{sec:2D:self-energy}
\begin{figure}
\begin{tabularx}{\columnwidth}{lXcX}
(a) &&
$\Sigma_\lambda=$
\begin{tikzpicture}[baseline={(0,-0.07)}]
\draw[dashed] (0,0) -- (0,1);
\fill (0,0) circle (0.07) node[anchor=north]{$\sqrt{\alpha_\lambda}$};
\tikzset{shift={(0,1)}}
\tstar{0.03}{0.1}{5}{200}{fill=black}
\tikzset{shift={(1.5,-1)}}
\end{tikzpicture}
$+$
\begin{tikzpicture}[baseline={(0,-0.07)}]
\draw [-{Stealth[length=3mm, width=2mm]}] (0.75,0) -- (0.87,0) node[anchor=south]{$\lambda$};
\draw (0,0.025) -- (1.5,0.025);
\draw (0,-0.025) -- (1.5,-0.025);
\fill (0,0) circle (0.07) node[anchor=north]{$\sqrt{\alpha_\lambda}$};
\fill (1.5,0) circle (0.07)
node[anchor=north]{$\sqrt{\alpha_\lambda}$};
\draw[dashed] (0,0) -- +(0.75,1);
\draw[dashed] (1.5,0) -- +(-0.75,1);
\tikzset{shift={(0.75,1)}}
\tstar{0.03}{0.1}{5}{20}{fill=black}
\end{tikzpicture}
$+$
\begin{tikzpicture}[baseline={(0,-0.07)}]
\draw [-{Stealth[length=3mm, width=2mm]}] (0.75,0) -- (0.87,0) node[anchor=south]{$\bar\lambda$};
\draw (0,0.025) -- (1.5,0.025);
\draw (0,-0.025) -- (1.5,-0.025);
\fill (-0.07,-0.07) rectangle (0.07,0.07)
node[xshift=-5pt, yshift=-2pt,anchor=north]{$\sqrt{\beta}$};
\fill (1.43,-0.07) rectangle (1.57,0.07)
node[xshift=-5pt, yshift=-2pt,anchor=north]{$\sqrt{\beta}$};
\draw[dashed] (0,0) -- +(0.75,1);
\draw[dashed] (1.5,0) -- +(-0.75,1);
\tikzset{shift={(0.75,1)}}
\tstar{0.03}{0.1}{5}{20}{fill=black}
\end{tikzpicture}
&\\
\addlinespace[8pt]
(b)&&
$\Sigma_{\lambda \bar\lambda} = $
\begin{tikzpicture}[baseline={(0,-0.07)}]
\draw[dashed] (0,0) -- (0,1);
\fill (-0.07,-0.07) rectangle (0.07,0.07) node[xshift=-5pt, yshift=-2pt,anchor=north]{$\sqrt{\beta}$};
\tikzset{shift={(0,1)}}
\tstar{0.03}{0.1}{5}{200}{fill=black}
\tikzset{shift={(1.5,-1)}}
\end{tikzpicture}
$+$
\begin{tikzpicture}[baseline={(0,-0.07)}]
\draw [-{Stealth[length=3mm, width=2mm]}] (0.75,0) -- (0.87,0) node[anchor=south]{$\lambda$};
\draw (0,0.025) -- (1.5,0.025);
\draw (0,-0.025) -- (1.5,-0.025);
\fill (0,0) circle (0.07)
node[anchor=north]{$\sqrt{\alpha_\lambda}$};
\fill (1.43,-0.07) rectangle (1.57,0.07)
node[xshift=-5pt, yshift=-2pt,anchor=north]{$\sqrt{\beta}$};
\draw[dashed] (0,0) -- +(0.75,1);
\draw[dashed] (1.5,0) -- +(-0.75,1);
\tikzset{shift={(0.75,1)}}
\tstar{0.03}{0.1}{5}{20}{fill=black}
\end{tikzpicture}
$+$
\begin{tikzpicture}[baseline={(0,-0.07)}]
\draw [-{Stealth[length=3mm, width=2mm]}] (0.75,0) -- (0.87,0) node[anchor=south]{$\bar\lambda$};
\draw (0,0.025) -- (1.5,0.025);
\draw (0,-0.025) -- (1.5,-0.025);
\fill (-0.07,-0.07) rectangle (0.07,0.07)
node[xshift=-5pt, yshift=-2pt,anchor=north]{$\sqrt{\beta}$};
\fill (1.5,0) circle (0.07)
node[anchor=north]{$\sqrt{\alpha_{\bar\lambda}}$};
\draw[dashed] (0,0) -- +(0.75,1);
\draw[dashed] (1.5,0) -- +(-0.75,1);
\tikzset{shift={(0.75,1)}}
\tstar{0.03}{0.1}{5}{20}{fill=black}
\end{tikzpicture}
&
\end{tabularx}
\caption{The irreducible diagrams contributing to the self-energy up to second order. Circular (quadratic) vertices denote intraband (interband) scattering events, stars the impurities. Panel (a) shows the diagonal contributions to the self-energy $\Sigma_{\lambda\lambda} \equiv \Sigma_\lambda$ whereas Panel (b) shows the off-diagonal contributions which are suppressed and hence neglected in the following.}
\label{fig:feynman}
\end{figure}
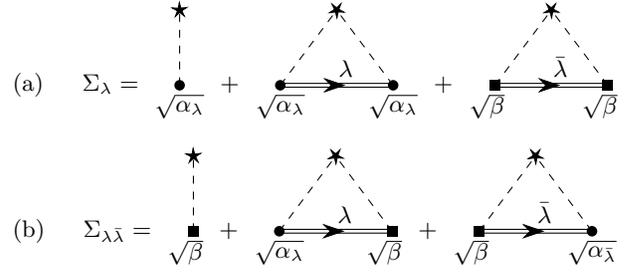

In general, impurities lead to spectral broadening of the LLs captured by the band-dependent Dingle temperatures $T_{D,\lambda}$ which are normally of the order of a few Kelvin. We will show in the following that the band-dependent self-energy can acquire oscillations with the frequencies associated with both Fermi surfaces.

To calculate the self-energy we use the self-consistent Born approximation (SCBA). A graphical representation of contributing irreducible diagrams is shown in Fig.~\ref{fig:feynman}. The first order contributions are scattering events on a single impurity. Already at this level it is obvious that the interband channels of the scattering vertex lead to a non-diagonal self-energy, see Fig.~\ref{fig:feynman}~(b). This is due to the fact that the impurities do not conserve the quantum number $\lambda$.

The central approximation we do in order to make analytic progress is to neglect the off-diagonal elements of the self-energy, $\Sigma_{\lambda \bar\lambda}=0$, which is at least in the limit $\nicefrac{T_{D,\lambda}}{|W_1-W_2|} \ll 1$ rigorously justified \cite{Raikh1994}. {\cred The main idea is that only the Landau levels in a range $T_D$ around the Fermi energy are relevant for
transitions. The off-diagonal elements couple Landau levels well split in energy, rendering their effect on the Green’s function irrelevant. Hence,} the effect of the first order diagrams can be incorporated by simply renormalizing $W_\lambda - n_\text{imp} U_0 \sqrt{\alpha_\lambda}\rightarrow W_\lambda$. The second order contributions are double scattering events on a single impurity. Due to the random and uniform distribution of the impurities the self-energy can be expressed as an integral
\begin{equation}
\Sigma_\lambda = n_\text{imp} \frac{U_0^2}{L_x L_y} \int \d^2 r \left(\alpha_\lambda G_\lambda(\v{r},\v{r},\xi) + \beta G_{\bar\lambda}(\v{r},\v{r},\xi) \right),
\label{eq:2D:self-energy_integral}
\end{equation}
with the full Green's function in real space defined by
\begin{equation}
G_\lambda(\v{r},\v{r}',E) = \sum_{l, k_x} \frac{\Psi^*_{\lambda, l, k_x}(\v{r}') \Psi_{\lambda, l, k_x}(\v{r})}{E-\epsilon_\lambda(l)-\Sigma_\lambda(E)}.
\label{eq:2D:Greensfunction_realspace}
\end{equation}
and the electron wave function given by
\begin{equation}
\Psi_{\lambda, l, k_x}(\v{r}) = \frac{\e^{\ii k_x x}}{\sqrt{L_x}} \psi_{l}(y-y_0).
\label{eq:2D:wavefct}
\end{equation}
Here, $\psi_{l}(y-y_0)$ are the Eigenfunctions of a harmonic oscillator located at $y_0 =  \nicefrac{k_x}{e B}$. An explicit calculation of the Green's function for $\v{r} = \v{r}'$ can now be carried out, see Appendix~\ref{app:2D:self-energy}, and it turns out to be spatially invariant
\begin{align} 
G_\lambda(\v{r},\v{r},\xi) = -\ii \frac{m_\lambda}{2}  \left(1+2\sum_{k=1}^\infty (-1)^k \e^{2 \pi \ii k (\xi_\lambda^\star + \ii \Gamma_\lambda(\xi))}\right).
\label{eq:2D:Greensfunction_result}
\end{align}
The evaluation of the spatial integral in \eqref{eq:2D:self-energy_integral} is therefore trivial. \eqref{eq:2D:Greensfunction_result} can be obtained from \eqref{eq:2D:Greensfunction_realspace} by using the summation over $k_x$ to integrate out the dependence on the wavefunction and transforming the sum over LLs $l$ by Poisson summation to a sum over harmonics.

Motivated by the fact that the Dingle temperature $T_{D,\lambda}$ is a measure of the total interactions in the system, we set $\pi T_{D,\lambda} = |\langle\Im \Sigma_\lambda\rangle|$. Additionally, we introduce $\tilde \alpha_\lambda$ and $\tilde \beta_\lambda$ as weights of the different contributions (consider them as renormalized values of $\alpha$ and $\beta$, see Appendix~\ref{app:2D:self-energy} for details) and for convenience an operator $\mac{A}$ with the property $\mac{A}x_\lambda = x_{\bar\lambda} \mac{A}$ and $\mac{A} = 1$ if it is all the way to the right.

In this compact notation, we obtain the self-consistent equations for the self-energy
\begin{widetext}
\begin{align}
|\Im \Sigma_\lambda(\xi)| &= \pi T_{D,\lambda} \left[1 + \left(\tilde\alpha_\lambda+\tilde\beta_\lambda \mathcal{A} \right)  \sum_{k=1}^\infty (-1)^k \cos\left(2 \pi k \xi^\star_\lambda \right) R_\lambda(\xi)^k \right] 
\label{eq:2D:self-energy_selfconsistent1} \\
\Re \Sigma_\lambda(\xi) &= \pi T_{D,\lambda} \left(\tilde\alpha_\lambda+\tilde\beta_\lambda \mathcal{A} \right)   \sum_{k=1}^\infty (-1)^k  \sin\left(2 \pi k \xi^\star_\lambda \right) R_\lambda(\xi)^k.
\label{eq:2D:self-energy_selfconsistent2}
\end{align}
\end{widetext}
Although \eqref{eq:2D:self-energy_selfconsistent1} and \eqref{eq:2D:self-energy_selfconsistent2} still need to be solved for $\Sigma_\lambda$, they already show an intriguing property of the self-energy -- the self-energy of an electron of type $\lambda$ oscillates not only with the basis frequency which is dictated by its own Fermi energy (the $\tilde\alpha_\lambda$ contribution) but also with a frequency associated with the respectively other Fermi surface (the $\tilde\beta_\lambda$ contribution). This contribution is solely a result of interband scattering, which provides the non-linear coupling. 
 
The self-consistent equations \eqref{eq:2D:self-energy_selfconsistent1} and (\ref{eq:2D:self-energy_selfconsistent2}) can be solved by iterative insertion in the strong damping limit $R_{D,\lambda} \ll 1$ which translates to $T_{D,\lambda} \gtrsim \omega_{c\lambda}$. However we argue that all results hold also qualitatively in the limit where $R_{D,\lambda}$ approaches 1. This seems reasonable since higher order effects are simply additive and therefore may change amplitude and phase factors of the oscillations but will not change the frequencies or their dependence on temperature. 

\subsubsection{Quantum Oscillations and temperature smearing}
We now solve the self-consistent equations for the self-energy up to second order in $R_{D,\lambda}$. Then we insert our findings in the conductivity kernel \eqref{eq:2D:cond_result} keeping only terms up to the second harmonic, i.e. second order in $R_{D,\lambda}$. 

At second order the interference of the various oscillating quantities leads not only to a change of the non-universal amplitudes but also to new frequencies. More precisely, the oscillations of the conductivity kernel interfere with the oscillating Dingle factor, an oscillating prefactor and an oscillating contribution to the chemical potential. Since these oscillations all originate from oscillations of the self-energy, there is an interference of $F_1$ with $F_2$, which appears as oscillations of the form $\cos(2\pi[\xi_1\pm\xi_2])$.

Various frequencies contribute to the conductivity but they all follow the generic description of \eqref{eq:condKernel_genericForm} such that the evaluation of the integrals can be carried out. The evaluation of the convolution with $n_F'$, see \eqref{eq:convolution}, leads to a LK temperature damping following \eqref{eq:cond_f}, details are given in the Appendix~\ref{app:temperature convolution}. 

Neglecting all non-oscillatory terms, the final result in 2D then reads 
\begin{align}
\frac{\sigma_{xx}}{\sigma_0} =& \sum_\lambda A_{1F\lambda} \cos \left(2 \pi\frac{\mu+W_\lambda}{\omega_\lambda}\right)  R_{D,\lambda} R_\text{LK}\left(2 \pi^2 \frac{T}{\omega_\lambda} \right) \nonumber \\
&+ \sum_\lambda A_{2F\lambda} \cos \left(4 \pi\frac{\mu+W_\lambda}{\omega_\lambda}\right)  R_{D,\lambda}^2 R_\text{LK}\left(4 \pi^2 \frac{T}{\omega_\lambda} \right)
\nonumber \\
&+ A_{+} \cos \left(2 \pi\frac{\mu+W_+}{\omega_+}\right)  R_{D,1} R_{D,2} R_\text{LK}\left(2 \pi^2 \frac{T}{\omega_+}\right)
\nonumber \\
&+ A_{-} \cos \left(2 \pi\frac{\mu+W_-}{\omega_-}\right)  R_{D,1} R_{D,2} R_\text{LK}\left(2 \pi^2 \frac{T}{\omega_-}\right)
\label{eq:2D:cond_final}
\end{align}
where $\omega_\pm^{-1}=\omega_1^{-1} \pm \omega_2^{-1}$, $\frac{\mu+W_\pm}{\omega_\pm} = \frac{\mu+W_1}{\omega_1} \pm \frac{\mu+W_2}{\omega_2}$ and the non-universal amplitudes $A = A(\mu)$ are evaluated at the Fermi energy/chemical potential and are presented in \eqref{app:2D:A_0} to \eqref{app:2D:A last} of the Appendix. As expected the sum and difference frequency are only present for nonzero interband scattering $A_\pm \propto \beta$. 

We note that our results are in agreement with previous calculation on magneto  inter-subband oscillations for 2DEGs~\cite{Polyanovsky1988, Coleridge1990, Raikh1994}. So far, our calculations are mainly a generalization from previous derivations as we considered different effective masses. In the remainder of this manuscript we establish that similar results hold also true in isotropic 3D metals and for relativistic dispersions.

\begin{figure*}
\includegraphics[scale=1]{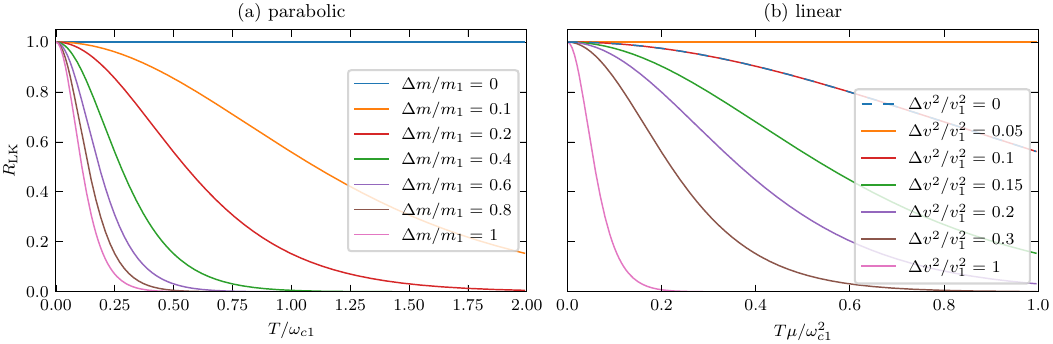}
\caption{Damping factor of the difference frequency for parabolic bands with various values of $\Delta m = m_1-m_2$, panel (a) and relativistic bands with various values of $\Delta v^2 = v_2^2-v_1^2$, panel (b). For parabolic bands the damping factor becomes independent of temperature for equal effective masses $\Delta m = 0$. In the relativistic case we set for this plot $W_2=0$ and $\nicefrac{W}{\mu}=\nicefrac{1}{20}$, such that the damping factor remains slightly temperature dependent for equal Fermi velocities $\Delta v^2 = 0$. It becomes however independent of temperature for $\Delta v^2 = \nicefrac{W_1}{\mu}$.}
\label{fig3}
\end{figure*}

\subsection{Three dimensions}
\label{sec:3B}
The 2D calculations can be easily generalized to a generic three dimensional metal. The cross section of a Fermi surface of a 3D metal may be complicated such that for simplicity we assume that there exist only two electron or hole pockets between which electrons can be scattered by a channel $\beta$. Note that within the second order SCBA this treatment can be easily generalized to multiple electron/hole pockets. We expand around the Fermi pockets assuming a local, rotational symmetric, quadratic dispersion such that momenta should be understood as crystal momenta with respect to the center of the Fermi pockets. The Hamiltonian is again given by \eqref{eq:2D_hamiltonian} but with three dimensional momenta.

The dispersion of the LLs is now continuous in $z$-direction $\epsilon_\lambda(l,k_z) = \omega_{c\lambda} (l+\nicefrac{1}{2}) + \nicefrac{k_z^2}{2m_\lambda}  - W_\lambda$. We can now show that the effect of the $k_z$-dependence of the dispersion only results in an additional phase of the QOs, similar to the canonical LK theory, see e.g. Ref.~\cite{Shoenberg1984}.

In 3D the wavefunction includes now the additional factor $\e^{\ii k_z z} /\sqrt{L_z}$ but $G(\v{r},\v{r},\xi)$ remains spatially invariant. We can simply change our equations from the 2D case to describe the 3D model by transforming $W_\lambda \rightarrow W_\lambda-\nicefrac{k_z^2}{2m_\lambda}$ and integrate over $k_z$ momenta. The resulting integrals are of the form 
\begin{equation}
\int_{-\sqrt{\xi}}^{\sqrt{\xi}} \d x \left(\xi-x^2 \right)^n \e^{2 \pi \ii\left(\xi-x^2\right)} = J_n(\xi).
\label{eq:3D_integral}
\end{equation}
Using the usual approximation of a large Fermi energy $\xi \gg 1$ it can be shown that $J_n(\xi) = \frac{\xi^n}{\sqrt{2}}  \e^{2\pi \ii \xi-\ii \nicefrac{\pi}{4}}$, see Appendix \ref{app:evaluation of J_n}, where in general the phase consists of a variety of contributions making it non-universal. Details of the full calculation are presented in the Appendix~\ref{app:3D}.

The final result for the conductivity in 3D resembles the 2D one, \eqref{eq:2D:cond_final},  with different non-universal amplitudes $A$ and an additional phase $\phi$ inside the $\cos$ terms, see \eqref{app:3D:cond_final} for the full expression of the QO or Table~\ref{tab:summary} for a summary. We can conclude that sum and difference frequencies of two Fermi pockets are observeable as long as the Fermi pockets are sufficiently strongly coupled.

\section{Relativistic dispersions}
\subsection{Double Weyl model in 2D}
A material class, which has all ingredients for temperature stable difference frequencies, are multifold fermion systems~\cite{manes2012existence,Bernevig2012,bradlyn2016beyond}. In these, bands which are parallel within a large region of the Brillouin zone arise out of symmetry arguments, e.g. in representatives of space group 198 \cite{manes2012existence}. However, the quasiparticles near the Fermi energy are often relativistic, hence possess a linear energy-momentum dispersion. In this section we show that sum and difference frequencies also emerge in this situation and follow the general form of \eqref{eq:cond_f}.

We first consider a model which consists of two Weyl cones labeled by $\lambda=1,2$ in 2D. The Hamiltonian can be written in the pseudospin basis, $\{\Phi\}$, and in the band basis, $\{\Psi\}$, as
\begin{align}
H_0 &= \sum_{\v{k},\lambda} \v{\Phi}_{\v{k},\lambda}^\dag \left(v_\lambda \v{\tau} \cdot \v{k} - W_\lambda \1 \right) \v{\Phi}_{\v{k},\lambda} \nonumber \\
&= \sum_{\v{k},\sigma=\pm,\lambda} \epsilon_{\lambda,\sigma}(\v{k}) \Psi_{\v{k},\sigma,\lambda}^\dag \Psi_{\v{k},\sigma,\lambda},
\label{eq:Weyl:Hamiltonian}
\end{align}
with $\epsilon_{\sigma,\lambda}(\v{k}) = \sigma v_\lambda|\v{k}|-W_\lambda$, $\tau_i$ are the Pauli matrices $i = x,y $ and the subband index $\sigma = \pm 1$ labels the two subbands of each Weyl cone. The two cones are shifted in energy by $W_1-W_2$, see Fig.~\ref{fig1} (b). The extremal cross sectional Fermi surface looks the same as for quadratic bands in 2D or 3D (for fixed $k_z$), see Fig.~\ref{fig1} (c). Note that we refer to the band structure as double Weyl cones, but it equivalently applies for all other linearly dispersive band structures, e.g. Dirac cones or band structures which can be effectively described by the double Weyl model around the Fermi energy. 

The LLs of relativistic electrons are not equidistant but follow $\epsilon_{\sigma,\lambda}(l) = \sigma \omega_{c \lambda} \sqrt{l}-W_\lambda$ where the cyclotron frequency $\omega_{c\lambda}=v_\lambda \sqrt{2 e B}$ now depends on the Fermi velocity $v_{\lambda}$ \cite{Castro2009}. Their wavefunctions are symmetric and antisymmetric superpositions of neighboring harmonic oscillator levels with different pseudospin, see \eqref{eq:Weyl:wavefunction}.

As above, we add impurities to the system by adding a scattering potential 
\begin{equation}
\mac{U} = \sum_{\v{r}} U(\v{r}) \Psi^\dag_{\v{r},\sigma,\lambda} \left(\Lambda_{\lambda, \lambda'} \otimes \delta_{\sigma,\sigma'} \right) \Psi_{\v{r},\sigma',\lambda'}
\end{equation}
to the Hamiltonian $H = H_0 + \mac{U}$. The scattering vertex $\Lambda \otimes \1$ (with $\Lambda$ from \eqref{eq:2D:scattering_vertex}) features intracone intraband scattering ($\alpha$--channels) and intercone scattering ($\beta$--channel) but does not have any intracone intersubband channels. The choice of a trivial intracone intersubband vertex is motivated by the graphene literature \cite{Peres2006} and simplifies the calculation of the self-energy. As before, our central assumption in the analytical derivation will be that the self-energy remains diagonal in the cone index $\lambda$ neglecting off-diagonal elements of the self-energy.

\subsubsection{Conductivity}
The calculation of the conductivity for Weyl cones differs from that  of parabolic bands. The velocity operator in the pseudospin basis reads simply $v_x = \tau_x$. However we perform our calculations in the band basis with a magnetic field. In this case $v_x$ couples different LLs as well as different bands. An evaluation of the trace in the conductivity kernel yields
\begin{align}
\hat \sigma_{xx}(E) =& \frac{e^2 N_\Phi}{\pi V} \sum_{l,\lambda} v_\lambda^2 \Im(G_{l,+,\lambda}+G_{l,-,\lambda}) \nonumber\\
&\times \Im(G_{l-1,+,\lambda}+G_{l-1,-,\lambda})
\end{align}
in agreement with e.g. Ref.~\cite{Peres2006}.

The sum over LLs can be transformed into a sum over harmonics by Poisson summation as before and we obtain
\begin{align}
\hat \sigma_{xx}(\xi) =& \frac{4e^2}{\pi}\sum_\lambda \xi_\lambda^* |\Gamma_\lambda|
\frac{\xi_\lambda^{*2}+\Gamma_\lambda^2}{1+16 \xi_\lambda^{*2} \Gamma_\lambda^2} \nonumber \\
&\times \left[1+2\sum_{k>0} \cos\left(2\pi k (\xi_\lambda^{*2}-\Gamma_\lambda^2)\right) R_\lambda^k\right].
\label{eq:Weyl:cond_kernel_full}
\end{align}
where now $R_\lambda(\xi) = \exp\left(-4\pi \xi_\lambda^* |\Gamma_\lambda|\right)$ in contrast to parabolic bands, compare to \eqref{eq:DampinfFac}. We note that the same conductivity kernel for a single Weyl cone has already been derived in Ref.~\cite{Gorbar2002, Gusynin2005} but we argue that our derivation of a more general form is considerably simpler.

\subsubsection{Self-energy}
The real space Green's function $G_{\lambda, \sigma} (\v{r},\v{r'},\xi)$ depends now additionally on the subband index $\sigma$. We can reuse \eqref{eq:2D:Greensfunction_realspace} keeping in mind the different dispersion of relativistic LLs. The wavefunction  
\begin{equation}
\Psi_{\sigma \lambda,l,k_x}(\v{r}) = \frac{\e^{\ii k_x x}}{\sqrt{2 L_x}}  \left(\psi_{l} (y-y_0) + \sigma \psi_{l-1} (y-y_0)\right) 
\label{eq:Weyl:wavefunction}
\end{equation}
 mixes different levels of the harmonic oscillator. However, the calculation of $G_{\lambda, \sigma} (\v{r},\v{r},\xi)$ works analogously. The crucial step is to perform first the summation over subbands $\sigma$ before transforming the sum over LLs to a sum over harmonics in order to be able to use complex contour integration for the integral. We then find analogously to above, the self-consistent equations for the self-energy
\begin{widetext}
\begin{align}
|\Im \Sigma_\lambda(\xi)|
=& \pi T_{D,\lambda}\left[1+ \left(\tilde\alpha_\lambda + \tilde \beta_\lambda \cal{A} \right) \sum_{k=1}^\infty \cos \left(2 \pi k \left(\xi^{\star 2}_{\lambda}-\Gamma_\lambda^2\right)\right)  R_{\lambda}^k(\xi) \right] \label{eq:Weyl:self-energy1} \\
\Re\Sigma_\lambda(\xi) 
=& \pi T_{D,\lambda}\left(\tilde\alpha_\lambda + \tilde \beta_\lambda \cal{A} \right) \sum_{k=1}^\infty \sin \left(2 \pi k \left(\xi^{\star 2}_{\lambda}-\Gamma_\lambda^2\right)\right)  R_{\lambda}^k(\xi) ,
\label{eq:Weyl:self-energy2}
\end{align}
\end{widetext}
where we used $\xi_\lambda^\star \gg \Gamma_\lambda$ outside the arguments of $\sin$ and $\cos$ and introduced the artificial Dingle--temperature as a prefactor. These equations should be seen as analogue to \eqref{eq:2D:self-energy_selfconsistent1} and (\ref{eq:2D:self-energy_selfconsistent2}). The main differences are all expected for Weyl systems, i.e. this is the quadratic dependence on $\xi$ in the arguments of $\cos$/$\sin$ and the explicit dependence of the Dingle factor on $\xi$.

\subsubsection{QOs and temperature smearing}
The expansion of the conductivity kernel and the evaluation of the temperature integral is analogous to the scenario with parabolic bands. In principle the $\Gamma_\lambda$ term inside the $\cos$ would lead to additional contributions to the amplitude, but these turn out to be suppressed with $\nicefrac{T_{D\lambda}}{\mu} \ll 1$. We neglect these minor changes in the frequency and note that they should become important at low fillings of the Weyl cone.

Omitting all non-oscillatory terms we finally obtain the main result for the conductivity as 
\begin{widetext}
\begin{align}
\frac{\sigma_{xx}}{\sigma_0} =& \sum_\lambda A_{1F\lambda} \cos \left(2 \pi\left[\frac{\mu+W_\lambda}{\omega_{c\lambda}}\right]^2\right)  R_{D,\lambda} R_\text{LK}\left(4 \pi^2 \frac{T(\mu+W_\lambda)}{\omega_{c\lambda}^2} \right) \nonumber \\
&+ \sum_\lambda A_{2F\lambda} \cos \left(4 \pi\left[\frac{\mu+W_\lambda}{\omega_{c\lambda}}\right]^2\right)  R_{D,\lambda}^2 R_\text{LK}\left(8 \pi^2 \frac{T(\mu+W_\lambda)}{\omega_{c\lambda}^2} \right)
\nonumber \\
&+ A_{+} \cos \left(2 \pi\left[\frac{(\mu+W_1)^2}{\omega_{c1}^2}+\frac{(\mu+W_2)^2}{\omega_{c2}^2}\right]\right)  R_{D,1} R_{D,2} R_\text{LK}\left(2 \pi^2 T \left[\frac{\mu+W_1}{\omega_{c1}^2}+\frac{\mu+W_2}{\omega_{c2}^2}\right]\right)
\nonumber \\
&+ A_{-} \cos \left(2 \pi\left[\frac{\mu+W_1}{\omega_{c1}}\right]^2-2\pi\left[\frac{\mu+W_2}{\omega_{c2}}\right]^2\right)  R_{D,1} R_{D,2} R_\text{LK}\left(2 \pi^2 T\left[\frac{\mu+W_1}{\omega_{c1}^2}-\frac{\mu+W_2}{\omega_{c2}^2}\right]\right)
\label{eq:Weyl:cond_final}
\end{align}
\end{widetext}
where the non-universal amplitudes $A$ are given in the Appendix \eqref{eq:app:Weyl:amplitude_first} to (\ref{eq:app:Weyl:amplitude_last}). Although \eqref{eq:Weyl:cond_final} seems complicated it can be understood equivalently to \eqref{eq:2D:cond_final}. The first and second line are the first and second harmonics of the basis frequencies $F_\lambda$ which are dictated by the geometry of the Fermi surface. The sum and difference frequencies of the basis frequencies, line three and four, are also of second order in the Dingle factor. The temperature dependence of the amplitudes follows our main result \eqref{eq:cond_f}.

Note that for relativistic dispersions the Fermi surface area depends  quadratically on the chemical potential. Hence, the scale for the exponential temperature decay is set by $\frac{\mu+W_\lambda}{v_\lambda^2}$ instead of $m_\lambda$ \cite{Kueppersbusch2017,Castro2009}.

\subsection{Three dimensions}
The Hamiltonian of the double Weyl model \eqref{eq:Weyl:Hamiltonian} can be easily generalized to three dimensions by including the Pauli--$z$ matrix $\tau^z$. The LLs are then continuous $\epsilon_{\sigma,\lambda}(l,k_z) = \sigma \sqrt{v_\lambda^2 k_z^2+w_{c\lambda}^2 l}-W_\lambda$. In 3D the wavefunction is more involved then before, since the prefactors of $\psi_l$ and $\psi_{l-1}$ in \eqref{eq:Weyl:wavefunction} depend now additionally on $l$, $z$ and $k_z$. The nested form of the dispersion and the non-trivial prefactors of the wavefunction make analytic calculations cumbersome. Nevertheless we can use a similar argument as in sec.~\ref{sec:3B}: Using the relation $\epsilon_{\sigma,\lambda}(l,k_z)=\epsilon_{\sigma,\lambda}\left(l+\frac{\nu_\lambda^2}{\omega_{c\lambda}^2} k_z^2\right)$ the $k_z$-dependence reduces to a real shift of the pole in \eqref{eq:app:Weyl:greensfct}. Hence, the terms $\xi_\lambda^{\star^2}-\Gamma_\lambda^2$ transform to $\xi_\lambda^{\star^2}-\Gamma_\lambda^2 - \nu_\lambda^2 k_z^2$, whereas terms $\xi_\lambda^\star \Gamma_\lambda$ which are the imaginary shifts of the poles remain the same. The appearing integrals are then of the form $J_n(\xi^2)$. Therefore, we conclude that an extension from 2D to 3D has for relativistic bands the same effect on the conductivity as for parabolic bands --- only additional non-universal phases shift the QOs but the overall phenomenology remains unchanged.

\begin{table*}
\begin{tabular}{l|l||ll|ll}
&&quadratic dispersion && linear dispersion&\\\hline
&&canonical & sum/difference & canonical & sum/difference \\
\hline \hline
&$f(\mu)$  &$2\pi k \frac{\mu + W}{\omega_c}$&  $2\pi \left(\frac{\mu + W_1}{\omega_{c1}} \pm \frac{\mu + W_2}{\omega_{c2}}\right)$&  
$2\pi k \left[\frac{\mu + W}{\omega_c}\right]^2$&  $2\pi \left(\left[\frac{\mu + W_1}{\omega_{c1}}\right]^2 \pm \left[\frac{\mu + W_2}{\omega_{c2}}\right]^2\right)$\\
2D& $\chi$ &$2\pi^2 k \frac{T}{\omega_c}$& $2\pi^2 \frac{T}{\omega_{c\pm}}$&
$4\pi^2 k T \frac{\mu + W}{\omega_c^2}$& $4\pi^2 T \left(\frac{\mu + W_1}{\omega_{c1}^2} \pm\frac{\mu + W_2}{\omega_{c2}^2}\right)$\\
&Dingle    &$2\pi^2 k \frac{T_D}{\omega_c}$& $2\pi^2 \left(\frac{T_{D1}}{\omega_{c1}} + \frac{T_{D2}}{\omega_{c2}}\right)$&
$4\pi^2 k T_D\frac{\mu+W}{\omega_c^2}$& $4\pi^2 \left(T_{D1}\frac{\mu+W_1}{\omega_{c1}^2} + T_{D2}\frac{\mu+W_2}{\omega_{c2}^2}\right)$\\
\hline
&$f(\mu)$  &$2\pi k \frac{\mu + W}{\omega_c} + \phi$&  $2\pi \left(\frac{\mu + W_1}{\omega_{c1}} \pm \frac{\mu + W_2}{\omega_{c2}}\right) + \phi_\pm$&  
$2\pi k \left[\frac{\mu + W}{\omega_c}\right]^2+\phi$&  $2\pi \left(\left[\frac{\mu + W_1}{\omega_{c1}}\right]^2 \pm \left[\frac{\mu + W_2}{\omega_{c2}}\right]^2\right) +\phi_\pm$ \\
3D& $\chi$ &$2\pi^2 k \frac{T}{\omega_c}$& $2\pi^2 \frac{T}{\omega_{c\pm}}$&$4\pi^2 k T \frac{\mu + W}{\omega_c^2}$& $4\pi^2 T \left(\frac{\mu + W_1}{\omega_{c1}^2} \pm\frac{\mu + W_2}{\omega_{c2}^2}\right)$\\
&Dingle    &$2\pi^2 k \frac{T_D}{\omega_c}$& $2\pi^2 \left(\frac{T_{D1}}{\omega_{c1}} + \frac{T_{D2}}{\omega_{c2}}\right)$&$4\pi^2 k T_D\frac{\mu+W}{\omega_c^2}$& $4\pi^2 \left(T_{D1}\frac{\mu+W_1}{\omega_{c1}^2} + T_{D2}\frac{\mu+W_2}{\omega_{c2}^2}\right)$\\
\end{tabular}
\caption{Overview of results. We summarize the behavior for the argument of the oscillations, $f$ as defined in Eq.~\ref{eq:cond_f}, the argument of the LK temperature damping factor, $\chi$ as defined in Eq.~\ref{eq:LK1}, and the argument of the impurity damping factor, e.g. the Dingle factor as defined in Eq.~\ref{eq:DingleDamping}. For the canonical basis frequency we have included the behavior of higher harmonics with integers $k$.}
\label{tab:summary}
\end{table*}

 \section{de Haas--van Alphen effect}
The main objective of our work is to establish  difference frequency QOs as a generic phenomenon of multiband metals. We focused on the Shubnikov--de Haas effect, i.e. QOs of the conductivity. In this section we comment on the behaviour of the de Haas--van Alphen effect, i.e. the QOs of quantities derived from the thermodynamic potential.

The main result of this section is that to 2nd order in $R_D$ the difference frequency is absent in the density of states and therefore in the de Haas--van Alphen effect, whereas the sum frequency remains observable. This applies equivalently for all higher order combination frequencies. 

We evaluate the density of states per unit area 
\begin{equation}
\rho(E) = -\frac{1}{\pi L_x L_y} \tr_{l,k_x,\lambda} \left[ \Im G(E) \right]
\label{eq:dHvA:DOS_def}
\end{equation}
from the imaginary part of the retarded, impurity averaged Green's function. The result for the density of states for parabolic bands in 2D
\begin{equation}
\rho(E) = \sum_\lambda \frac{m_\lambda}{2\pi} \left(1+ 2\sum_{k=1}^\infty (-1)^k \cos(2\pi k \xi_\lambda^\star) R_\lambda(\xi)^k\right)
\label{eq:dHvA:DOS_res}
\end{equation}
is derived in appendix~\ref{sec:app:E1} and should be seen as the de Haas--van Alphen analogue of \eqref{eq:2D:cond_result}. Hence, we continue with the same expansion up to second order in the Dingle factor as for the conductivity. The major difference in the result is that the  difference frequency term cancels exactly in the expansion because the self-energy enters in the density of states only over the damping factor $R_\lambda$ and $\xi_\lambda^\star$ and not via any prefactors like the scattering time for the conductivity.

The grand canonical potential {\cred of a fermionic system is well known (see e.g. Ref.~\cite{Shoenberg1984}) 
\begin{equation}
\Omega = -T\int_{-\infty}^\infty \d E \log\left(1+\e^{-(E-\mu)/T}\right) \rho(E) .
\end{equation}}
To keep the analogy to the derivation of the conductivity, we {\cred rewrite the integration, using integration by parts twice, into a convolution involving} the zero temperature grand canonical potential
\begin{equation}
\hat\Omega(\mu) = \int_{{\cred -\infty}}^\infty \d E \Theta(\mu-E) \rho(E) {\cred( E- \mu)}
\label{eq:dHvA:Omega0}
\end{equation}
and include temperature in the same way as for the conductivity \eqref{eq:convolution}
\begin{equation}
\Omega = \int_{-\infty}^\infty \d \varepsilon [-n_F'(\varepsilon)] \hat \Omega(\mu+\varepsilon).
\end{equation}

Note that the integration in \eqref{eq:dHvA:Omega0} has no effect on the oscillations up to a $\nicefrac{\pi}{2}$ phase change. The result for the oscillating part of the thermodynamic potential in 2D is
\begin{align}
\Omega =& \sum_\lambda A_{1F\lambda} \cos \left(2 \pi\frac{\mu+W_\lambda}{\omega_\lambda}\right)  R_{D,\lambda} R_\text{LK}\left(2 \pi^2 \frac{T}{\omega_\lambda} \right) \nonumber \nonumber \\
&+ \sum_\lambda A_{2F\lambda} \cos \left(4 \pi\frac{\mu+W_\lambda}{\omega_\lambda}\right)  R_{D,\lambda}^2 R_\text{LK}\left(4 \pi^2 \frac{T}{\omega_\lambda} \right) \nonumber
\nonumber \\
&+ A_{+} \cos \left(2 \pi\frac{\mu+W_+}{\omega_+}\right)  R_{D,1} R_{D,2}
R_\text{LK}\left(2 \pi^2 \frac{T}{\omega_+} \right).
\label{eq:dHvA:omega_res}
\end{align}
and the amplitudes are given in \eqref{eq:app:dHvA:2D:As1}-\eqref{eq:app:dHvA:2D:As-1}. Strikingly, only the difference frequency is absent but the other frequencies show the same behavior as in the conductivity.

The exact cancellation of the difference frequency is not an artifact of the model. It remains also valid in 3D and in linear band structures, see appendix~\ref{sec:app:E2}. For relativistic dispersions a difference frequency appears, however its amplitude is negligible compared to the sum frequency.

We like to point out that a difference frequency can be generated in the de Haas--van Alphen effect by higher order scattering processes. Going to the third order of the SCBA, i.~e. three scattering events on the same impurity, difference and sum frequency are already generated at the level of the self-energy, but again the amplitude is expected to be strongly reduced.

\section{Discussion and Materials}
We have shown in detail how to compute QOs of the conductivity for  two-band models with parabolic and relativistic dispersions in 2 and 3D. Remarkably, we find that a non-linear coupling of bands --- studied in terms of interband impurity scattering --- leads to the emergence of new sum and difference frequencies. Their amplitudes are damped with the sum and difference of the temperature scales of their basis frequencies. Hence, a striking feature is that the difference does not acquire any temperature dependence at all if the effective masses of two parabolic bands are the same. For relativistic bands the point of absolute temperature stability is a fine-tuned one, depending on the relation between $v_\lambda$, $W_\lambda$ and $\mu$. For parallel linear bands, i.e. equal Fermi velocities $v_1 = v_2$, the difference frequency remains slightly temperature damped by $\frac{W_1-W_2}{v_1^2}$, providing an opportunity to experimentally distinguish relativistic and parabolic dispersion. In Tab.~\ref{tab:summary} we present a concise summary of our calculations. 

We conclude this section by discussing experimental requirements for observing temperature stable difference frequencies, and furthermore, point out possible candidate materials. The main ingredient for an appreciable amplitude of the difference frequency is, of course, an electronic band structure with multiple pockets whose quasiparticles have similar mass (or Fermi velocity). 
The main non-trivial requirement is a strong effective coupling $\beta$ between the bands. The exact strength of the interband scattering depends on the type of the impurity and on the microscopic details of the wavefunction. Therefore, we expect that \textit{ab-initio} calculations will be very helpful for estimating the inter- versus intraband impurity scattering strengths in suitable materials. 
In addition, the strength/density of intraband impurities influences strongly the Dingle temperatures. Our expansion in $R_{D, \lambda}$ does in principle require $T_{D, \lambda} \gtrsim \omega_{c\lambda}$, however we argue that our expansion also holds qualitatively for $R_{D, \lambda} \rightarrow 1$. Hence, in addition to an effective interband coupling, the observation of the difference frequency is mainly limited by the strength of the signals of the second harmonics of the basis frequencies which are similarly of second order in the Dingle factors. We argue that strong signals from the higher harmonics are a good indication that the difference frequency can be observed. These can be maximized by choosing a relatively clean system with $w_{c \lambda} \approx T_{D, \lambda}$. Finally, we note that the amplitudes $A_\pm$ in 3D are slightly suppressed for large frequencies.

Next, we discuss concrete material candidates. Intriguingly,  difference and sum frequencies have potentially been already  measured in various systems, but their existence has not been attributed to the present mechanism from interband coupling. For example, the 3D heavy-fermion superconductor CeCoIn$_5$ displays QO frequencies which are approximately the difference of two larger frequencies~\cite{Shishido2018}, and which persist when the material is doped with Nd~\cite{Klotz2018}. Similarly, the tritelluride NdTe$_3$ shows a frequency which is the difference of two basis frequencies to high accuracy~\cite{Dalgaard2020}.

In general, we expect materials with multifold fermion excitations to be prime candidates because they have parallel bands over large momentum space regions. For example, the topological semimetal PtGa shows a difference frequency and several other frequencies in the Fourier spectrum whose origins are unexplained~\cite{ShengXu2020}. Most strikingly, in accordance with our predictions a temperature stable difference frequency has been very recently reported for the topological semi-metal CoSi~\cite{Huber2023}.

Similarly, Shoenberg's classic book on QOs~\cite{Shoenberg1984} lists large number of materials displaying putative magnetic breakdown frequencies some of which could be difference frequencies. It would also be worthwhile to search within the recent class of square-net materials~\cite{klemenz2019topological} for unusual QOs which fall outside the scope of the standard LK theory.

Beyond the 2D systems studied previously in the context of {\it magneto intersubband oscillations} of 2DEGs~\cite{Coleridge1990,Leadley1992, Goran2009}, systems like bilayer graphene~\cite{mccann2013electronic,mccann2006landau} show all the required properties of the band structure but to our knowledge no difference frequency has been reported. This might be related to ineffective interband scattering but we expect that the controlled introduction of selected impurities could do the trick, which is again an oustanding task for {\it ab-initio} modelling. We note that recently a difference frequency has been reported in twisted bilayer graphene where interband scattering is induced by imperfections of the Moire pattern \cite{Phinney2021_strong}.

Another general expectation is that band splitting is not only induced by interlayer tunneling but also via spin orbit coupling, e.g. for Rashba surface states~\cite{sunko2017maximal}, which could result in temperature stable difference frequencies. An observation thereof would turn difference frequency QOs into a very precise tool for determining the energy scales of spin orbit induced band splitting.

\section{Outlook}
We have shown how non-linear interband coupling influences the Fourier spectrum of QOs up to second order in the Dingle factors with the emergence of a new difference frequency stable in temperature, see Fig.~\ref{fig1}~(d). A natural next question is which other higher order effects can emerge in the QO spectrum? Based on our calculations, we expect that any integer, linear combination of the basis frequencies can appear as well as the interference of higher harmonics. These higher order QOs will be damped with the Dingle factors of the involved frequencies and acquire a temperature smearing according to \eqref{eq:cond_f}.

For the calculation of the self-energy we have used the SCBA. Going beyond this, the full SCBA would take into account multiple scattering events on a single impurity by including products of real-space Green's functions. Hence, sum and difference frequencies should already appear at the level of the self-energy but we expect that the resulting QOs are qualitatively similar and behave in the same way as the higher harmonics discussed above. 

For the non-linear interband coupling we have concentrated on the effect of impurities. However, the coupling of bands can also be the result of interactions. Clearly, the Coulomb interaction between electrons is not limited to electrons of the same band. In an expansion of the self-energy the Fock diagrams resemble those of the impurity scattering shown in Fig.~\ref{fig:feynman}~(a) and are expected to lead to similar effects. In that context, the effect of Coulomb interactions on the de Haas--van Alphen effect of quasi-2D systems has been recently studied in Ref.~\cite{Allocca2021}, which also finds a difference frequency albeit with a strong temperature dependence. Similarly,  interaction mediated fluctuations have recently been shown to enhance de Haas--van Alphen QOs in insulators~\cite{allocca2022quantum,allocca2023fluctuation}. In general, a detailed study on the interplay of interband coupling from impurities and interactions for thermodynamic and transport QOs remains a formidable task for the future.  

The unambiguous observation of a new difference frequency in QOs is  exciting by itself~\cite{Huber2023}. In addition, because of its temperature stability it can be turned into a versatile tool, for example, for studying the temperature dependence of the Dingle temperature, for quantifying interband scattering strengths or band splitting mechanisms like spin orbit coupling. In this way difference frequency QO measurements may detect temperature dependent changes of material properties which are otherwise impossible to observe with the strongly damped canonical QOs. 

In conclusion, difference frequency QOs are a qualitatively new phenomenon beyond the known magnetic breakdown scenarios. Despite the long history of QO research we expect further surprises in the future.

\begin{acknowledgments}
We acknowledge a related experimental collaboration and discussions with  N.~Huber, M.~Wilde and C.~Pfleiderer. We thank N.~R.~Cooper for helpful discussions. V.~L. thanks S.~Birnkammer for helpful discussions.
V.~L. acknowledges support from the Studienstiftung des deutschen Volkes. J.~K. acknowledges support from the Imperial-TUM flagship partnership. The research is part of the Munich Quantum Valley, which is supported by the Bavarian state government with funds from the Hightech Agenda Bayern Plus.
\end{acknowledgments}

\section*{Code availability}
The symbolic calculations related to this paper are available on Zenodo \cite{code} from the corresponding authors
upon reasonable request.

{\it Note added:}
After submission of our work, Ref.~\cite{Alisultanov2023} appeared which also considered the effect of linear dispersions on magneto intersubband oscillations in layered quasi-2D systems. Their conclusions agree with ours for the 2D results (bulk 3D dispersions are not addressed in their work).

\appendix

\section{Evaluation of the temperature convolution}
\label{app:temperature convolution}
 In this section we derive the temperature dependence of the conductivity out of the zero temperature conductivity kernel. For a generic conductivity kernels we obtain in leading order a temperature dependence of LK type. The calculation below applies for any oscillating conductivity kernel, being applicable for quadratic and linear dispersion in any dimension.

Starting from the generic form the conductivity kernel \eqref{eq:condKernel_genericForm}, we expand the functions $f(E)$ and $g(E)$ around the Fermi energy $\mu$ motivated by the form of the integral in \eqref{eq:convolution}. We truncate the expansion of $f(E)$ around $E = \mu + \varepsilon$ at linear order, higher orders can be taken into account systematically. The convolution of the conductivity may then be written as
\begin{align}
\sigma =& \Re \e^{\ii f(\mu)} \int_{-\infty}^{\infty} \d \varepsilon [-n_F'(\varepsilon)] \e^{\ii f'(\mu) \varepsilon} \nonumber\\
&\times \sum_{n=0}^\infty \frac{g^{(n)}(\nicefrac{\mu}{\omega_c})}{n!} \frac{\d^n g}{\d \xi^n}\biggr|_{\nicefrac{\mu}{\omega_c}} \left(\frac{\varepsilon}{\omega_c}\right)^n 
\\
=& \Re \e^{\ii f(\mu)} \sum_{n=0}^\infty \frac{\d^n g}{\d \xi^n}\biggr|_{\nicefrac{\mu}{\omega_c}} \left(\frac{T}{\omega_c}\right)^n I_n (f'(\mu) T)
\label{eq:App:cond_all orders}
\end{align}
where we have introduced the integral
\begin{equation}
I_n (a) = \frac{1}{n!} \int_{-\infty}^{\infty} \d x  \frac{\e^{ \\i a x}}{\left(\e^{x/2}+\e^{-x/2}\right)^2} x^n 
\end{equation}
which can be evaluated exactly.

\subsection{Evaluation of $I_n$}
Obviously $\Re I_{2n+1}(a)=\Im I_{2n}(a)=0$ due to antisymmetry of the integrands. In the following we will show that the integrals are given by derivatives of the Lifshitz-Kosevich damping factor
\begin{align}
\Re I_{2n} (a) =& \frac{(-1)^n}{(2n)!} \frac{\d^{2n}}{\d \lambda^{2n}} \left(\frac{\lambda}{a}\right)^{2n} \frac{1}{\lambda}  R_\text{LK}\left(\pi\frac{a}{\lambda}\right) \biggr|_{\lambda=1} \label{eq:App:I_n_res}\\
\Im I_{2n+1} (a) =& \frac{(-1)^n}{(2n+1)!} \frac{\d^{2n}}{\d \lambda^{2n}} \left(\frac{\lambda}{a}\right)^{2n+1} \frac{1}{\lambda^2} R_\text{LK}\left(\pi\frac{a}{\lambda}\right)
\nonumber\\ &\times
\left[2n-1+ R_\text{LK}\left(\pi\frac{a}{\lambda}\right)  \cosh\left(\pi\frac{a}{\lambda}\right)\right]  \biggr|_{\lambda=1}.
\end{align}
The calculation below shows how the expression for $\Re I_{2n}$ can be obtained, the calculation for $\Im I_{2n+1}$ is analogous. Using the Geometric series we rewrite the exponential factors in the denominator of the integral. However,  the geometric series holds strictly speaking only true for $x>0$, hence we take the integration boundaries  from $\epsilon \to 0$ to $\infty$ and perform the limit in the end to ensure proper convergence. For the following calculation we set $n$ to be even:
\begin{widetext}
\begin{align}
n! \Re I_n(a) =& 2 \int_\epsilon^\infty \d x  \frac{\cos(a x)}{\left(\e^{x/2}+\e^{-x/2}\right)^2} x^n \nonumber\\
=& -2 \sum_{k=1}^\infty (-1)^k k \int_\epsilon^\infty \d x \e^{-kx} \cos(ax) x^n \nonumber\\
=& -2 \sum_{k=1}^\infty (-1)^k k^{1-n} (-1)^n \frac{\d^n}{\d \lambda^n} \int_\epsilon^\infty \d x \e^{-\lambda kx} \cos(ax)\biggr|_{\lambda=1} \nonumber\\
=& -\frac{\d^n}{\d \lambda^n}\sum_{k=-\infty}^\infty (-1)^k   \frac{\lambda k^{2-n}}{a^2+\lambda^2 k^2} \e^{-\epsilon \lambda |k|} \biggr|_{\lambda=1} \nonumber\\
=& \frac{\d^n}{\d \lambda^n} \sum_{z^*=\pm \ii \frac{a}{\lambda}} \text{Res} \left( \frac{\lambda z^{2-n}}{a^2+\lambda^2 z^2} \frac{\pi}{\sin(\pi z)} \e^{-\epsilon \lambda z}, z=z^* \right) \biggr|_{\lambda=1} \nonumber\\
=& (-1)^\frac{n}{2} \frac{\d^n}{\d \lambda^n} \left(\frac{\lambda}{a}\right)^n \frac{\pi \frac{a}{\lambda^2}}{\sinh\left(\pi \frac{a}{\lambda}\right)} \biggr|_{\lambda=1}
\end{align}
\end{widetext}
Note that we have used the Feynman trick in line 3, that the summand for $k=0$ vanishes for all values of $n$ (line 4) and the residue formula for summation in line 5.

\subsection{Interpretation of the temperature dependence}
From \eqref{eq:App:cond_all orders} and \eqref{eq:App:I_n_res} it is obvious that the lowest order of the expansion in $g$ (i.e. $n=0$) leads to a temperature dependence of LK type. Higher order corrections ($n>0$) to the temperature dependence are typically suppressed by the functional form of $g$ --- for $g$ being e.g. a polynomial the $n$-th derivative of $g$ is suppressed with $\xi^{-n}$.

\section{Supplement for 2D parabolic dispersions}
This section provides several details to the calculations performed in the sec.~\ref{sec:3A}. First details on the calculation of the self-energy are explained, second the used scheme to expand the conductivity kernel to second or even higher order is demonstrated.

\subsection{Calculation of the self-energy}
\label{app:2D:self-energy}
In order to evaluate the self-energy $\Sigma_\lambda$ \eqref{eq:2D:self-energy_integral} we need to determine the real space Green's function for $\v{r}=\v{r}'$ from \eqref{eq:2D:Greensfunction_realspace}. We use the summation over $k_x$-momenta to integrate out the $y$-dependencies and use Poisson summation \eqref{eq:poisson_sum}
\begin{align}
G_\lambda(\v{r},\v{r},E) &= \frac{1}{L_x} \sum_{l, k_x} \frac{\psi^*_{\lambda,l}(y-y_0) \psi_{\lambda,l}(y-y_0)}{E-\epsilon_\lambda(l)-\Sigma_\lambda(E)} \\
&=   \frac{1}{2 \pi} \sum_{l=0}^\infty \int_{-\infty}^{\infty} \d k_x \frac{|\psi_{\lambda,l}(y-y_0)|^2}{E-\epsilon_\lambda(l)-\Sigma_\lambda(E)} \\
&= \frac{e B}{2 \pi \omega_\lambda} \sum_{k=-\infty}^\infty (-1)^k \int_{\frac{1}{2}}^{\infty} \d u\frac{\e^{2 \pi \ii k u}}{\xi_\lambda^\star - u +\ii \Gamma_\lambda}.
\end{align}
The integral over $u$ can be evaluated for $k\neq 0$ by using $\xi_\lambda^\star\gg 1$ and complex contour integration:
\begin{align}
& \quad \int_{\frac{1}{2}}^{\infty} \d u\frac{\e^{2 \pi \ii k v}}{-\xi_\lambda^\star+ u-\ii \Gamma_\lambda(\xi)} \\
&\approx \e^{2 \pi \ii k \xi_\lambda^*} \int_{-\infty}^{\infty} \d u\frac{\e^{2 \pi \ii k u}}{u -\ii \Gamma_\lambda(\xi)} \\
&= 2\pi\ii \e^{2 \pi \ii k (\xi_\lambda^\star +\ii\Gamma_\lambda(\xi))}\Theta(k \Gamma_\lambda(\xi)) \text{sgn}(\Gamma_\lambda(\xi)) .
\end{align}
For $k=0$ the integral is divergent, due to the divergent sum over $l$ in \eqref{eq:2D:Greensfunction_realspace} since  we did not assume that our energy spectrum is bounded from above. It is however easy to see that the imaginary part of the integral is convergent
\begin{align}
\Im \int_{\frac{1}{2}-\xi_\lambda^\star}^{\infty} \frac{\d u}{u - \ii \Gamma_\lambda} \approx \pi \text{sgn}(\Gamma_\lambda).
\end{align}
For the real part of the integral we introduce an upper cut off $\Lambda_c$ and use that the integrand is odd
\begin{align}
\Re \int_{\frac{1}{2}-\xi_\lambda^\star}^{\Lambda_c} \frac{\d u}{u - \ii \Gamma_\lambda} = \frac{1}{2} \log \left(\frac{\Lambda_c^2+\Gamma_\lambda^2}{\left(\xi_\lambda^\star-\frac{1}{2}\right)^2+\Gamma_\lambda^2} \right).
\end{align}
Anticipating that $\Sigma_\lambda \sim T_{D,\lambda}$ which is of order of a few Kelvin, any physical oscillations of this formally divergent part are suppressed with at least $\frac{T_{D,\lambda}}{\mu} \ll 1$ and can be neglected. We absorb the non/weakly-oscillating real part in the chemical potential~\cite{Grigoriev2003,Thomas2008}, to obtain the oscillating Green's function \eqref{eq:2D:Greensfunction_result}.

The correct prefactors for the interband and intraband contribution are motivated from the fact that the Dingle temperature is a measure of the total interactions in the system $\pi T_{D,\lambda} = |\langle \Im \Sigma_\lambda\rangle|$. We introduce the band dependent total mass $M_\lambda = \alpha_\lambda m_\lambda + \beta m_{\bar{\lambda}}$ and the Dingle temperature $\pi T_{D,\lambda} = \frac{1}{2}n_\text{imp} U_0^2 M_\lambda$ leading to \eqref{eq:2D:Greensfunction_realspace}. To simplify the notation, we also set $\tilde \alpha_\lambda = 2\alpha_\lambda\frac{m_\lambda}{M_\lambda}$ and $\tilde \beta_\lambda = 2\beta \frac{m_{\bar\lambda}}{M_\lambda}$.

We note at this point that several similar integrals need to be evaluated throughout this manuscript, e.g. in the derivation of \eqref{eq:2D:cond_result}. All of them can be evaluated in a similar fashion as above, the rest is however convergent unless stated differently.

\subsection{Expansion of the conductivity kernel}
In \eqref{eq:2D:cond_result} there are three terms that are oscillating with respect to the magnetic field: $\xi_\lambda^*$ taking into account the oscillating real part of the self-energy, the oscillations of $\Im \Sigma_\lambda$ leading effectively to an oscillating Dingle factor $R_\lambda(\xi)$ and to oscillations of the prefactor, and the intrinsic oscillations of the conductivity. Interestingly the oscillations of the Dingle factor and the real part of the self-energy cancel exactly for the difference frequency. We expand the conductivity up to second order in  $R_{D, \lambda} = R_\lambda\left(\nicefrac{\pi T_{D,\lambda}}{\omega_{c\lambda}}\right)$. There is no need to expand $\xi_\lambda^*$ if it appears outside the arguments of $\cos$ or $\sin$, since these second order contributions will be suppressed by $|T_{D,\lambda}/(\mu-W_\lambda)|\ll 1$. Therefore, $\Re \Sigma_\lambda$ only needs to be expanded up to first order, as it will only appear together with first order terms
\begin{align}
\Re \Sigma^{(1)}_\lambda(\xi) &= -\pi T_{D,\lambda} \left(\tilde\alpha_\lambda+\tilde\beta_\lambda\mathcal{A}\right) \sin\left(2 \pi \xi_\lambda\right) R_{D,\lambda}.
\end{align}
The other expanded quantities read
\begin{align}
R^{(2)}_\lambda(\xi) =& R_{D,\lambda}\left[1 + 2\pi\tau_\lambda \left(\tilde\alpha_\lambda+\tilde\beta_\lambda\mathcal{A}\right) \cos\left(2 \pi \xi_\lambda\right) R_{D,\lambda} \right]
\\
|\Gamma^{(2)}_\lambda(\xi)| =& \tau_{\lambda} \left[1 -  \left(\tilde\alpha_\lambda+\tilde\beta_\lambda\mathcal{A}\right) \cos\left(2 \pi \xi^\star_\lambda\right) R^{(2)}_\lambda(\xi) \right.\nonumber\\&
+\left.\left(\tilde\alpha_\lambda+\tilde\beta_\lambda\mathcal{A}\right)  \cos\left(4 \pi \xi_\lambda\right) R_{D,\lambda}^2
\right]
\\
\sigma_{xx}^{(2)}(\xi) =& \sigma_0 \sum_\lambda \frac{\xi_\lambda \Gamma^{(2)}_\lambda(\xi)}{1+4\left(\Gamma^{(2)}_\lambda(\xi)\right)^2}
\nonumber\\
&\times\left(1 - 2\cos \left(2\pi \xi^*_\lambda\right) R^{(2)}_\lambda(\xi)
+2 \cos \left(4\pi \xi_\lambda\right) R_{D,\lambda}^{2} \right)
\end{align}
where $\tau_\lambda = \frac{\pi T_{D,\lambda}}{\omega_\lambda}$ is the dimensionless Dingle temperature.

After an expansion, done with \textit{Mathematica} \cite{code} where we collect terms with the same frequency we find for the conductivity kernel 
\begin{align}
\frac{\sigma_{xx}(\xi)}{\sigma_0} =& A_0(\xi)  + \sum_\lambda \cos \left(2 \pi \xi_\lambda\right)R_{D,\lambda} A_{1F\lambda}(\xi)
\nonumber\\
&+ \sum_\lambda \cos \left(4 \pi \xi_\lambda\right) R_{D,\lambda}^2 A_{2F\lambda}(\xi)
\nonumber \\
&+ \cos \left(2 \pi\left[\xi_1+\xi_2 \right]\right) R_{D,1} R_{D,2} A_+(\xi)
\nonumber\\
&+ \cos \left(2 \pi\left[\xi_1-\xi_2 \right]\right) R_{D,1} R_{D,2} A_-(\xi).
\end{align}
The amplitudes read
\begin{widetext}
\begin{align}
A_0(\xi) =& 
\sum_\lambda \frac{\tau_\lambda \xi_\lambda}{1+4\tau_\lambda^2} + \frac{R_{D,\lambda}^2 \xi_\lambda}{\left(1+4\tau_\lambda^2\right)^3} \left(\tilde{\alpha }_{\lambda}\left[\tau_{\lambda }-16 \tau_{\lambda }^5\right] +\tilde{\alpha }_{\lambda }^2\left[-6 \tau_{\lambda}^3+8 \tau_{\lambda}^5\right] \right)
\nonumber\\
&+\frac{R_{D,\lambda}^2 \xi_{\bar\lambda} \tilde\beta_{\bar\lambda}^2}{\left(1+4\tau_{\bar\lambda}^2\right)^3} \left(-6 \tau_{\bar{\lambda }}^3+8 \tau_{\bar{\lambda }}^5\right)
\label{app:2D:A_0}
\\
A_{1F\lambda}(\xi) =&
-\frac{2\tau_{\lambda }\xi_\lambda}{4 \tau_{\lambda}^2+1}-\tilde\alpha_\lambda\left(\frac{\tau_{\lambda }\xi_\lambda}{4 \tau _{\lambda}^2+1}
-\frac{8 \tau_{\lambda}^3\xi_\lambda}{\left(4 \tau _{\lambda }^2+1\right)^2} \right)
- \tilde\beta _{\bar{\lambda }} \left(\frac{ \tau _{\bar{\lambda}} \xi_{\bar{\lambda }}}{4 \tau _{\bar{\lambda }}^2+1}-\frac{8 \tau_{\bar{\lambda }}^3 \xi_{\bar{\lambda }}}{\left(4 \tau _{\bar{\lambda }}^2+1\right)^2}\right)
\\
A_{2F\lambda}(\xi) =&
\frac{\xi _{\lambda }}{\left(4 \tau _{\lambda }^2+1\right)^3} \Big(2 \tau _{\lambda }+16 \tau _{\lambda }^3+32 \tau _{\lambda }^5 \nonumber \\
&
\left.+\tilde\alpha_\lambda\left[2 \tau _{\lambda }-4 \pi  \tau _{\lambda }^2-32 \pi  \tau _{\lambda }^4-32 \tau _{\lambda }^5-64 \pi  \tau _{\lambda }^6\right]+\tilde{\alpha }_{\lambda }^2\left[-2 \pi  \tau _{\lambda }^2-6 \tau _{\lambda }^3+8 \tau _{\lambda }^5+32 \pi  \tau _{\lambda }^6\right] \right)
\nonumber\\
&+
\frac{\tilde{\beta }_{\bar{\lambda }} \xi _{\bar{\lambda }}}{\left(1+4 \tau _{\bar{\lambda }}^2\right)^3} \left( \tau _{\bar{\lambda }}-16 \tau _{\bar{\lambda }}^5 +\tilde{\alpha }_{\lambda }\left[-2 \pi  \tau _{\lambda } \tau _{\bar{\lambda }}+32 \pi  \tau _{\lambda } \tau _{\bar{\lambda }}^5\right]+\tilde{\beta }_{\bar{\lambda }}\left[-6 \tau_{\bar{\lambda }}^3+8 \tau _{\bar{\lambda }}^5 \right] \right)
\\
A_+(\xi) =& 
\sum_\lambda \frac{\xi_{\lambda}\tilde{\beta}_{\lambda}}{\left(1+4 \tau_{\lambda }^2\right)^3} \left(\tau _{\lambda }-4 \pi  \tau _{\lambda }^2-32 \pi  \tau_{\lambda }^4-16 \tau _{\lambda}^5-64 \pi  \tau_{\lambda}^6
\right. \nonumber \\ &
+ \tilde{\alpha}_{\lambda}\left[-2 \pi  \tau_{\lambda }^2-12 \tau_{\lambda}^3+16 \tau_{\lambda }^5+32 \pi  \tau_{\lambda}^6\right]
+\left.
\tilde{\beta }_{\bar{\lambda }}\left[-2 \pi  \tau_{\lambda } \tau_{\bar{\lambda }}+32 \pi  \tau_{\lambda }^5 \tau_{\bar{\lambda }}\right] \right)
\\
A_-(\xi) =& 
\sum_\lambda \frac{\xi_{\lambda}\tilde{\beta}_{\lambda}}{\left(1+4 \tau_{\lambda }^2\right)^3} \left( \tau_{\lambda }-16 \tau_{\lambda}^5+\tilde{\alpha }_{\lambda}\left(-12 \tau_{\lambda }^3+16 \tau_{\lambda }^5\right) \right)
\label{app:2D:A last}
\end{align}
\end{widetext}
where $A_0$ constitutes the non-oscillating contributions which we state here for the sake of completeness, but we will drop it in all other calculations.

\section{Calculation for 3D parabolic dispersions}
\label{app:3D}

\subsection{Evaluation of the integral $J_n$}
\label{app:evaluation of J_n}
In order to evaluate the integral appearing in 3D calculations, see  \eqref{eq:3D_integral}, we would like to extend the integration boundaries to $\pm \infty$. However doing this directly would lead to a divergent integral for $n>0$ among other problems. We solve this problem by first using Feynman's integral trick and then extending the integration boundaries to $\pm \infty$
\begin{align}
J_n(\xi) =& (2 \pi \ii)^{-n} \frac{\d^n}{\d \lambda^n} \biggr|_{\lambda=1} \int_{-\sqrt{\xi}}^{\sqrt{\xi}} \d x \e^{2 \pi \ii\lambda\left(\xi-x^2\right)} \\
=& (2 \pi \ii)^{-n} \frac{\d^n}{\d \lambda^n}\biggr|_{\lambda=1} \frac{1}{\sqrt{2\lambda}}  \e^{2 \pi \ii \lambda \xi-\ii \nicefrac{\pi}{4}} \\
=& \frac{\xi^n}{\sqrt{2}} \e^{2 \pi \ii \xi-\ii \nicefrac{\pi}{4}}
\end{align}
where we used that $\xi\gg 1$ to obtain the last line.  To check the correctness of this calculation we compared the result to a numerical evaluation of \eqref{eq:3D_integral}.

In practice the $k_z$-integral will lead to an additional phase of $\nicefrac{\pi}{4}$, a suppression of oscillations with $\nicefrac{1}{\sqrt{\xi_\lambda}}$ as already predicted in Ref.~\cite{Polyanovsky1988} and a suppression of higher harmonics with $\nicefrac{1}{\sqrt{k}}$. These effects are clearly visible in \eqref{app:3D:cond} and \eqref{app:3D:self-energy}.

\subsection{Conductivity}
Starting from \eqref{eq:2D:cond_result} and carrying out the integral over $k_z$-momenta leads to 
\begin{align}
\hat \sigma_{xx} =& \frac{\sigma_0}{\pi \ell_B} \sum_{\lambda} \frac{\xi_\lambda^\star |\Gamma_\lambda(\xi)|}{1+4\Gamma_\lambda(\xi)^2} \times \left(\frac{2\sqrt{2}}{3} \sqrt{\xi_\lambda^\star} \right.
\nonumber\\
&\left.
+ \sum_{k=1}^\infty \frac{(-1)^k}{\sqrt{k}} \cos\left(2\pi k \xi_\lambda^\star-\frac{\pi}{4}\right) R_\lambda(\xi)^k\right)
\label{app:3D:cond}
\end{align}
where $\ell_B = \frac{1}{\sqrt{e B}}$ is the magnetic length scale. The integral for the first summand is evaluated exactly. Note that the self-energy does not depend on $k_z$ because we integrate that out. 

\subsection{Self-energy}
We start from \eqref{eq:2D:Greensfunction_result} to evaluate the integral and introduce the weights
$\tilde \alpha_\lambda = \frac{\alpha_\lambda m_\lambda \sqrt{2 m_\lambda E_\lambda}}{\alpha_\lambda m_\lambda \sqrt{2 m_\lambda E_\lambda}+\beta m_{\bar\lambda} \sqrt{2 m_{\bar\lambda} E_{\bar\lambda}}}$ and $\tilde \beta_\lambda = 1- \tilde \alpha_\lambda$. Since oscillations in $\xi_\lambda^\star$ are suppressed by $\nicefrac{1}{\xi_\lambda^\star}$ if they appear outside of $\cos$ or $\sin$ terms, we set $\xi_\lambda^\star = \xi_\lambda$ in these. Then we obtain the self-consistent equation for the self-energy
\begin{align}
\Sigma_\lambda =& -\ii \pi T_{D,\lambda}\biggr(1+ \left[\tilde \alpha_\lambda + \tilde \beta_\lambda \mac{A}\right] \frac{1}{\sqrt{2 \xi_\lambda}} 
\nonumber \\
&\times \sum_{k=1}^\infty \frac{(-1)^k}{\sqrt{k}} \e^{2\pi \ii k \xi_\lambda^\star - \ii \pi/4} R_\lambda(\xi)^k\biggr).
\label{app:3D:self-energy}
\end{align}
\eqref{app:3D:self-energy} is the analog of \eqref{eq:2D:self-energy_selfconsistent1} and \eqref{eq:2D:self-energy_selfconsistent2}. However, the QOs in the self-energy come in 3D with an additional small prefactor $\nicefrac{1}{\sqrt{\xi_\lambda}}$.

\subsection{QOs in 3D}
The expansion is done in the same manner as in 2D. However in 3D two types of second harmonics appear: The intrinsic second harmonic of $\Im \Sigma_\lambda$ and $\hat \sigma_{xx}$ with phase $\nicefrac{\pi}{4}$ and one resulting from interference with a phase $\nicefrac{\pi}{2}$. The later one is suppressed with $\nicefrac{1}{\sqrt{\xi_\lambda}}$ with respect the to other and hence neglected in the following.

The oscillating part of the conductivity reads 
\begin{widetext}
\begin{align}
\frac{\sigma_{xx}}{\sigma_0} =& \sum_\lambda A_{1F\lambda} \cos \left(2 \pi\frac{\mu+W_\lambda}{\omega_\lambda}-\frac{\pi}{4}\right)  R_{D,\lambda} R_\text{LK}\left(2 \pi^2 \frac{T}{\omega_\lambda} \right) 
+ \sum_\lambda A_{2F\lambda} \cos \left(4 \pi\frac{\mu+W_\lambda}{\omega_\lambda}-\frac{\pi}{4}\right)  R_{D,\lambda}^2 R_\text{LK}\left(4 \pi^2 \frac{T}{\omega_\lambda} \right)
\nonumber \\
&+ A_{+} \cos \left(2 \pi\frac{\mu+W_+}{\omega_+}-\frac{\pi}{2}\right)  R_{D,1} R_{D,2} R_\text{LK}\left(2 \pi^2 \frac{T}{\omega_+}\right)
+ A_{-} \cos \left(2 \pi\frac{\mu+W_-}{\omega_-}\right)  R_{D,1} R_{D,2} R_\text{LK}\left(2 \pi^2 \frac{T}{\omega_-}\right)
\label{app:3D:cond_final}
\end{align}
with the amplitudes \cite{code}
\begin{align}
A_{1F\lambda}(\xi) =&\frac{1}{\pi \ell_B} \left(  -\frac{\xi _{\lambda } \tau _{\lambda }}{1+4 \tau _{\lambda }^2}+\frac{16 \xi _{\lambda } \tau _{\lambda }^3 \tilde{\alpha }_{\lambda }}{3 \left(1+4 \tau _{\lambda }^2\right)^2}-\frac{2 \xi
   _{\lambda } \tau _{\lambda } \tilde{\alpha }_{\lambda }}{3 \left(1+4 \tau _{\lambda }^2\right)}+\frac{16 \xi _{\bar{\lambda}}^{\frac{3}{2}} \tau _{\bar{\lambda }}^3 \tilde{\beta }_{\bar{\lambda
   }}}{3 \sqrt{\xi _{\lambda }} \left(1+4 \tau _{\bar{\lambda }}^2\right)^2}-\frac{2 \xi _{\bar{\lambda }}^{\frac{3}{2}} \tau _{\bar{\lambda }} \tilde{\beta }_{\bar{\lambda }}}{3 \sqrt{\xi _{\lambda
   }} \left(1+4 \tau _{\bar{\lambda }}^2\right)}\right)
\\
A_{2F\lambda}(\xi) =& \frac{1}{\pi \ell_B} \left(\frac{\xi _{\lambda } \tau _{\lambda }}{\sqrt{2} \left(1+4 \tau _{\lambda }^2\right)}-\frac{8 \sqrt{2} \xi _{\lambda } \tau _{\lambda }^3 \tilde{\alpha }_{\lambda }}{3 \left(1+4 \tau _{\lambda
   }^2\right)^2}+\frac{\sqrt{2} \xi _{\lambda } \tau _{\lambda } \tilde{\alpha }_{\lambda }}{3 \left(1+4 \tau _{\lambda }^2\right)}-\frac{8 \sqrt{2} \xi _{\bar{\lambda }}^{\frac{3}{2}} \tau
   _{\bar{\lambda }}^3 \tilde{\beta }_{\bar{\lambda }}}{3 \sqrt{\xi _{\lambda }} \left(1+4 \tau _{\bar{\lambda }}^2\right)^2}+\frac{\sqrt{2} \xi _{\bar{\lambda }}^{\frac{3}{2}} \tau _{\bar{\lambda
   }} \tilde{\beta }_{\bar{\lambda }}}{3 \sqrt{\xi _{\lambda }} \left(1+4 \tau _{\bar{\lambda }}^2\right)} \right)
\nonumber\\
\\
A_+(\xi) =& \frac{1}{\pi \ell_B} \sum_{\lambda} \frac{\xi
   _{\lambda } \tilde{\beta }_{\lambda } }{\sqrt{\xi_{\bar{\lambda }}}
   \left(1+4 \tau _{\lambda}^2\right)}\left(\frac{\tau _{\lambda }}{2\sqrt{2}} - \sqrt{2} \pi  \tau _{\lambda}^2 - \frac{2\sqrt{2}}{3}  \pi  \tau_{\lambda }^2 \tilde{\alpha }_{\lambda } - \frac{2\sqrt{2}}{3}  \pi  \tau_{\lambda } \tau_{\bar{\lambda }} \tilde{\beta }_{\bar{\lambda }}\right)
   \nonumber \\ 
   &+\frac{\xi _{\lambda } \tilde{\beta }_{\lambda }
   }{\sqrt{\xi
   _{\bar{\lambda }}} \left(1+4 \tau _{\lambda }^2\right)^2} \left(-2 \sqrt{2} \tau _{\lambda }^3-4\sqrt{2} \tau _{\lambda }^3 \tilde{\alpha }_{\lambda } + \frac{16}{3} \sqrt{2} \pi  \tau _{\lambda }^4 \tilde{\alpha }_{\lambda } + \frac{16}{3} \sqrt{2} \pi  \tau _{\lambda }^3 \tau _{\bar{\lambda }} \tilde{\beta }_{\bar{\lambda }}\right)
   + \frac{64 \sqrt{2} \xi _{\lambda } \tau _{\lambda }^5 \tilde{\alpha }_{\lambda } \tilde{\beta }_{\lambda }}{3 \sqrt{\xi _{\bar{\lambda }}} \left(1+4 \tau _{\lambda }^2\right)^3}
\\
A_-(\xi) =& \frac{1}{\pi \ell_B} \sum_{\lambda} \frac{\xi
   _{\lambda } \tilde{\beta }_{\lambda } }{\sqrt{\xi_{\bar{\lambda }}}
   \left(1+4 \tau _{\lambda }^2\right)} \frac{\tau _{\lambda }}{2 \sqrt{2}}
   +\frac{\xi _{\lambda } \tilde{\beta }_{\lambda }
   }{\sqrt{\xi
   _{\bar{\lambda }}} \left(1+4 \tau_{\lambda}^2\right)^2} \left(-2 \sqrt{2} \tau_{\lambda}^3-4 \sqrt{2} \tau_{\lambda}^3 \tilde{\alpha }_{\lambda} \right)
   + \frac{64 \sqrt{2} \xi _{\lambda } \tau _{\lambda }^5 \tilde{\alpha }_{\lambda } \tilde{\beta }_{\lambda }}{3 \sqrt{\xi _{\bar{\lambda }}} \left(1+4 \tau _{\lambda }^2\right)^3}
\end{align}
\end{widetext}
which are evaluated at the chemical potential $A=A(\mu)$. Note that for 3D $\mac{O}(A_\pm) = \sqrt{\xi}$ whereas $\mac{O}(A_{2F\lambda}) = \xi$. This makes sum and difference frequencies more difficult to observe in 3D systems. 

\section{Supplement for 2D double Weyl model}
\subsection{Self-energy}
The Green's function in real space reads
\begin{equation}
G_{\sigma \lambda} (\v{r},\v{r}',\xi) = \frac{1}{\omega_\lambda} \sum_{l,k_x} \frac{\Psi_{\sigma \lambda,l,k_x}(\v{r})\Psi_{\sigma \lambda,l,k_x}(\v{r}')^*}{\xi_\lambda^\star-\sigma \sqrt{l}+\ii\Gamma_{\lambda}(\xi)}
\end{equation}
with the wavefunction given in \eqref{eq:Weyl:wavefunction}. For $\v{r}=\v{r}'$ we can sum out the wave function such that the real space Green's function is spatially invariant
\begin{align}
G_{\sigma\lambda}(\v{r},\v{r},\xi) =& \frac{e B}{4 \pi} \sum_{l} \frac{1}{E + W_\lambda-\sigma \omega_\lambda\sqrt{l} -\Sigma_{\lambda}} 
\nonumber \\
&\times
\int \d y_0 (|\psi_l(y-y_0)|^2+|\psi_{l-1}(y-y_0)|^2) \nonumber \\
=& -\frac{\sigma e B}{2 \pi \omega_\lambda} \sum_{k} \int_0^\infty \d l\frac{\e^{2\pi \ii k l}}{\sqrt{l} -\sigma \xi^*_{\lambda}-\sigma\ii \Gamma_{\lambda}} 
\end{align}
The self-energy can be easily calculated, since the Green's function does not depend on $\v{r}$ anymore and not on $\sigma$ 
\begin{align}
\Sigma_{\lambda}(\xi) = n_\text{imp} U_0^2 L_y L_x \sum_{\sigma}\left(\alpha_\lambda G_{\sigma\lambda}(\xi)+\beta G_{\sigma\bar\lambda}(\xi)\right)
\end{align}
The crucial part is the calculation of the term 
\begin{align}
\sum_{\sigma} G_{\sigma\lambda}(\xi) =& -\frac{e B}{2 \pi \omega_\lambda}
 \sum_{k}\int_0^\infty \d l \sum_{\sigma} \frac{\sigma\e^{2\pi \ii k l}}{\sqrt{l} -\sigma \xi^*_{\lambda}-\sigma\ii \Gamma_{\lambda}}
 \nonumber\\
 =&
 -\frac{e B}{2 \pi \omega_\lambda}
 \sum_{k}\int_0^\infty \d l \frac{2\e^{2\pi \ii k l}\left(\xi^*_{\lambda}+\ii \Gamma_{\lambda}\right)}{l -\left(\xi^*_{\lambda}+\ii \Gamma_{\lambda}\right)^2}
 \label{eq:app:Weyl:greensfct}
\end{align}
where we have used the Poisson summation formula to transform the sum over LLs to a sum over harmonics times an integral which we can compute with complex contour integration. We use again $\xi_\lambda^\star \gg 1$ such that the lower integration boundary can be shifted to $-\infty$. For $k=0$ the real part of this integral is divergent but the integrand is antisymmetric and will be set to zero. The imaginary part can be calculated explicitly
\begin{align}
\int_0^\infty \frac{\d l}{l -\left(\xi^*_{\lambda}+\ii \Gamma_{\lambda}\right)^2}
&=
\int_{-\infty}^\infty\d l \frac{l+\ii 2\xi^*_{\lambda}\Gamma_{\lambda}}{l^2 +4\left(\xi^*_{\lambda}\Gamma_{\lambda}\right)^2} \nonumber\\
&= \ii \pi \sgn(\Gamma_\lambda).
\end{align}
For the higher harmonics $k\neq 0$ we use complex contour integration
\begin{align}
&\sum_{k\neq0} \int_0^\infty \d l \frac{\e^{2\pi \ii k l}}{l -\left(\xi^*_{\lambda}+\ii \Gamma_{\lambda}\right)^2}
= \nonumber\\
& 2 \pi \ii \sgn(\Gamma_\lambda) \sum_{k>0}  \e^{2\pi \ii k \sgn(\Gamma_\lambda) \left(\xi^*_{\lambda}+\ii \Gamma_{\lambda}\right)^2}
\end{align}
in order to find in total
\begin{align}
\sum_{\sigma} G_{\sigma\lambda}(\xi) =& -\ii \frac{e B}{\omega_\lambda} \sgn(\Gamma_\lambda) \left(\xi^*_{\lambda}+\ii \Gamma_\lambda\right) \nonumber\\
&\times \left[1+2 \sum_{k>0}  \e^{2\pi \ii k \sgn(\Gamma_\lambda) \left(\xi^*_{\lambda}+\ii \Gamma_\lambda\right)^2}\right].
\end{align}
At this point it is useful to make several approximations to simplify the remaining calculations. First, we replace $|\langle \Im \Sigma_\lambda\rangle| = \omega_{c \lambda} |\langle \Gamma_\lambda\rangle|$ by the empirical Dingle temperature $\pi T_{D,\lambda}$. Note that $T_{D,\lambda}$ does not depend on the magnetic field as expected. Since the self-energy is of the order of the Dingle temperature which is only a few Kelvin we may use the approximations $\xi_\lambda^\star \approx \xi_\lambda$ and $\xi_\lambda^\star \gg \Gamma_\lambda$. This can however only be used outside the argument of the exponential terms. We then obtain the self-consistent \eqref{eq:Weyl:self-energy1} and \eqref{eq:Weyl:self-energy2}

\subsection{Amplitudes}
The conductivity kernel is expanded analogously to the parabolic case \cite{code}. The amplitudes read
\begin{widetext}
\begin{align}
A_{1F\lambda}(\xi) =& \frac{2 \xi _{\lambda }^3 \tau _{\lambda }}{1+16 \xi _{\lambda }^2 \tau _{\lambda }^2}
+\tilde{\alpha }_{\lambda } \left(\frac{\xi _{\lambda }^3 \tau _{\lambda }}{1+16 \xi _{\lambda }^2 \tau _{\lambda }^2} - \frac{32 \xi _{\lambda }^5 \tau _{\lambda }^3}{\left(1+16 \xi_{\lambda }^2 \tau _{\lambda }^2\right)^2}\right)
+\tilde{\beta }_{\bar{\lambda}} \left(\frac{\xi _{\bar{\lambda }}^3 \tau _{\bar{\lambda }}}{1+16 \xi _{\bar{\lambda }}^2 \tau _{\bar{\lambda }}^2} - \frac{32 \xi_{\bar{\lambda}}^5 \tau_{\bar{\lambda }}^3}{\left(1+16 \xi _{\bar{\lambda }}^2 \tau _{\bar{\lambda }}^2\right)^2}\right)
\label{eq:app:Weyl:amplitude_first}
\\
A_{2F\lambda}(\xi) =& 
\frac{2 \xi _{\lambda }^3 \tau _{\lambda }}{1+16 \xi _{\lambda }^2 \tau _{\lambda }^2}
   +\tilde{\alpha }_{\lambda } \left(\frac{2 \xi _{\lambda }^3 \tau _{\lambda }}{1+16 \xi _{\lambda }^2 \tau _{\lambda }^2}-\frac{64 \xi _{\lambda }^5 \tau _{\lambda }^3}{\left(1+16 \xi _{\lambda }^2 \tau _{\lambda }^2\right)^2}-\frac{8
   \pi  \xi _{\lambda }^4 \tau _{\lambda }^2}{1+16 \xi _{\lambda }^2 \tau _{\lambda }^2} -\frac{4 \pi \tilde\beta_{\bar \lambda} \xi_\lambda \tau_\lambda \xi _{\bar{\lambda }}^3 \tau _{\bar{\lambda}}\left(1-16 \xi_{\bar{\lambda }}^2 \tau_{\bar{\lambda
   }}^2\right)}{\left(1+16 \xi_{\bar{\lambda }}^2 \tau_{\bar{\lambda
   }}^2\right)^2}\right)
\nonumber\\
&-4\tilde{\alpha }_{\lambda}^2\xi _{\lambda }^4 \tau_{\lambda}^2 \frac{\pi+6 \xi_{\lambda } \tau_{\lambda }-32\xi_{\lambda}^3 \tau_{\lambda}^3-256\pi \xi_{\lambda}^4 \tau_{\lambda}^4}{\left(1+16 \xi _{\lambda }^2 \tau _{\lambda
   }^2\right)^3}+\tilde{\beta }_{\bar{\lambda }} \xi _{\bar{\lambda }}^3 \tau_{\bar{\lambda }} \frac{1-16 \xi _{\bar{\lambda }}^2 \tau _{\bar{\lambda}}^2}{\left(1+16 \xi _{\bar{\lambda }}^2 \tau_{\bar{\lambda }}^2\right)^2}-8 \tilde{\beta }_{\bar{\lambda }}^2 \xi _{\bar{\lambda }}^5 \tau_{\bar{\lambda }}^3 \frac{3-16 \xi _{\bar{\lambda }}^2 \tau _{\bar{\lambda
   }}^2}{\left(1+16 \xi _{\bar{\lambda }}^2 \tau_{\bar{\lambda }}^2\right)^3}
\\   
A_{+}(\xi) =& \sum_\lambda \tilde{\beta }_{\lambda } \left[\frac{\xi _{\lambda }^3 \tau _{\lambda }}{1+16 \xi _{\lambda }^2 \tau _{\lambda }^2}
-\frac{32 \xi _{\lambda }^5 \tau _{\lambda }^3}{\left(1+16 \xi _{\lambda }^2 \tau _{\lambda }^2\right)^2}- \frac{8 \pi  \xi _{\lambda }^4 \tau _{\lambda }^2}{1+16 \xi _{\lambda }^2 \tau _{\lambda }^2}\right.
\nonumber\\ & 
+ \tilde \alpha_\lambda \left(\frac{1024 \xi _{\lambda }^7 \tau
   _{\lambda }^5}{\left(1+16 \xi _{\lambda }^2 \tau _{\lambda }^2\right)^3}-\frac{48 \xi _{\lambda }^5 \tau_{\lambda}^3}{\left(1+16 \xi _{\lambda }^2 \tau_{\lambda }^2\right)^2} + \frac{128 \pi  \xi_{\lambda }^6 \tau _{\lambda }^4}{\left(1+16 \xi _{\lambda }^2 \tau _{\lambda }^2\right)^2} - \frac{4 \pi  \xi _{\lambda }^4 \tau _{\lambda }^2}{1+16 \xi _{\lambda }^2 \tau _{\lambda }^2}\right)
\nonumber\\
&\left.+\tilde{\beta }_{\bar{\lambda }} \left(\frac{128 \pi  \xi _{\lambda }^5 \xi _{\bar{\lambda }} \tau _{\lambda }^3 \tau _{\bar{\lambda }}}{\left(1+16 \xi _{\lambda }^2 \tau _{\lambda }^2\right)^2}-\frac{4 \pi  \xi _{\lambda }^3 \xi _{\bar{\lambda }} \tau _{\lambda }
   \tau _{\bar{\lambda }}}{1+16 \xi _{\lambda }^2 \tau _{\lambda }^2}\right)\right]
\\   
A_{-}(\xi) =&   \sum_\lambda \tilde{\beta}_{\lambda} \left[\frac{\xi _{\lambda }^3 \tau _{\lambda }}{1+16 \xi_{\lambda}^2 \tau_{\lambda}^2} -\frac{32 \xi_{\lambda}^5 \tau_{\lambda}^3}{\left(1+16 \xi_{\lambda }^2 \tau_{\lambda }^2\right)^2} + \tilde{\alpha}_{\lambda} \left(\frac{1024 \xi _{\lambda }^7
   \tau _{\lambda }^5}{\left(1+16 \xi _{\lambda }^2 \tau _{\lambda }^2\right){}^3}-\frac{48 \xi _{\lambda }^5 \tau _{\lambda }^3}{\left(1+16 \xi _{\lambda }^2 \tau _{\lambda }^2\right)^2}\right) \right]
\label{eq:app:Weyl:amplitude_last}
\end{align}
\end{widetext}

\section{Supplement for the de Haas--van Alphen effect}
\subsection{Calculation for 2D parabolic bands}
\label{sec:app:E1}
We evaluate the density of states \eqref{eq:dHvA:DOS_def} and show how to obtain \eqref{eq:dHvA:DOS_res}.
\begin{align}
\rho(E) =& \frac{1}{\pi L_x L_y } \sum_{k_x,l,\lambda} \frac{\Gamma_\lambda/\omega_\lambda}{\left(\xi_\lambda^\star-l-\frac{1}{2}\right)^2+\Gamma_\lambda^2}
\nonumber \\
=&\sum_{\lambda}\frac{N_\Phi \Gamma_\lambda}{\pi L_x L_y \omega_\lambda}\sum_l\frac{1}{\left(\xi_\lambda^\star-l-\frac{1}{2}\right)^2+\Gamma_\lambda^2}
\nonumber \\
=&\sum_{\lambda,k}\frac{m_\lambda \Gamma_\lambda}{2\pi^2} (-1)^k \e^{2\pi \ii k \xi_\lambda^\star} \int_{-\infty}^\infty \d l \frac{\e^{2\pi \ii k l}}{l^2+\Gamma_\lambda^2}
\nonumber \\
=&\sum_{\lambda,k}\frac{m_\lambda}{2\pi} \sgn \left(\Gamma_\lambda\right) (-1)^k \e^{2\pi \ii k \xi_\lambda^\star} R_{\lambda}^{|k|}(\xi)
\end{align}

The integrals to obtain the zero temperature thermodynamic potential \eqref{eq:dHvA:Omega0} are of the form
\begin{align}
\hat \Omega(\mu) =& \int_{{\cred 0}}^\infty \theta(\mu-E) \cos \left(2\pi\frac{E+W}{\omega_c}\right) {\cred( E- \mu)}
\nonumber \\
=& {\cred \frac{\omega_c^2}{4\pi^2}} {\cred \cos }\left(2\pi\frac{\mu+W}{\omega_c}\right).
\end{align}

Hence, the amplitudes for \eqref{eq:dHvA:omega_res} read \cite{code}
\begin{align}
A_{1F\lambda} &= -{\cred \frac{\omega_c}{8\pi^3 \ell_B^2}} \label{eq:app:dHvA:2D:As1}\\
A_{2F\lambda} &= {\cred \frac{\omega_c}{8\pi^3 \ell_B^2}} (1-2\pi \tilde \alpha_\lambda \tau_\lambda) \\
A_{+} &= -{\cred \frac{\omega_c}{4\pi^3 \ell_B^2 (m_1+m_2)}} (m_1 \tilde\beta_1\tau_1+m_2 \tilde\beta_2\tau_2).\label{eq:app:dHvA:2D:As-1}
\end{align}

\subsection{Calculation for relativistic dispersions in 2D}
\label{sec:app:E2}
The density of states reads 
\begin{equation}
\rho(E) = \sum_\lambda \frac{\xi_\lambda^\star}{\sqrt{2}\pi v_\lambda \ell_B} \left[1+2\sum_{k>0} \cos\left(2\pi k (\xi_\lambda^{\star 2}-\Gamma_\lambda^2)\right) R_\lambda^k\right].
\end{equation}
The additional prefactor of $\xi_\lambda^\star$ gives rise to the difference frequency. However the oscillating terms of the prefactor are small compared to the other oscillations. We take only the first non-vanishing order into account. After an expansion up to second order in the Dingle factor, we obtain
\begin{align}
\rho(E) =& \sum_\lambda A_{1F\lambda} \cos\left(2\pi \xi_\lambda^{2}\right) R_{D\lambda}(\xi)
\nonumber \\
&+\sum_\lambda A_{2F\lambda} \cos\left(4\pi \xi_\lambda^{2}\right) R_{D\lambda}(\xi)^2
\nonumber \\
&+ A_{+} \cos\left(2\pi \left[\xi_1^{2}+\xi_2^{2}\right]\right) R_{D1}(\xi) R_{D2}(\xi)
\nonumber \\
&+ A_{-} \sin\left(2\pi \left[\xi_1^{2}-\xi_2^{2}\right]\right) R_{D1}(\xi) R_{D2}(\xi) 
\end{align}
for the oscillating part of the density of states. The amplitudes read
\begin{align}
A_{1F\lambda} &= \frac{\sqrt{2} \xi_\lambda}{\pi v_\lambda \ell_B} \label{eq:app:dHvA:2D_Weyl:As1}\\
A_{2F\lambda} &= \frac{\sqrt{2} \xi_\lambda}{\pi v_\lambda \ell_B} -  \frac{4 \sqrt{2}}{\nu_\lambda \ell_B} \xi_\lambda^2 \tilde\alpha_\lambda\tau_\lambda\\
A_{+} &= -\frac{4\sqrt{2}}{\ell_B}\left(\frac{\xi_1^2 \tilde\beta_1\tau_1}{v_1}+ \frac{\xi_2^2 \tilde\beta_2\tau_2}{v_2}\right)
\\
A_{-} &= \frac{1}{\sqrt{2}\pi \ell_B}\left(\frac{\tilde\beta_1 \tau_1}{v_1}-\frac{\tilde\beta_2 \tau_2}{v_2}\right).
\label{eq:app:dHvA:2D_Weyl:As-1}
\end{align}
Note that the amplitude of the difference frequency is of order $\mathcal{O}(\nicefrac{A_-}{A_+}) \approx \nicefrac{1}{\xi^2}$ and cancels exactly for equal band parameters.

\bibliography{bib}

\begin{thebibliography}{70}%
\makeatletter
\providecommand \@ifxundefined [1]{%
 \@ifx{#1\undefined}
}%
\providecommand \@ifnum [1]{%
 \ifnum #1\expandafter \@firstoftwo
 \else \expandafter \@secondoftwo
 \fi
}%
\providecommand \@ifx [1]{%
 \ifx #1\expandafter \@firstoftwo
 \else \expandafter \@secondoftwo
 \fi
}%
\providecommand \natexlab [1]{#1}%
\providecommand \enquote  [1]{``#1''}%
\providecommand \bibnamefont  [1]{#1}%
\providecommand \bibfnamefont [1]{#1}%
\providecommand \citenamefont [1]{#1}%
\providecommand \href@noop [0]{\@secondoftwo}%
\providecommand \href [0]{\begingroup \@sanitize@url \@href}%
\providecommand \@href[1]{\@@startlink{#1}\@@href}%
\providecommand \@@href[1]{\endgroup#1\@@endlink}%
\providecommand \@sanitize@url [0]{\catcode `\\12\catcode `\$12\catcode
  `\&12\catcode `\#12\catcode `\^12\catcode `\_12\catcode `\%12\relax}%
\providecommand \@@startlink[1]{}%
\providecommand \@@endlink[0]{}%
\providecommand \url  [0]{\begingroup\@sanitize@url \@url }%
\providecommand \@url [1]{\endgroup\@href {#1}{\urlprefix }}%
\providecommand \urlprefix  [0]{URL }%
\providecommand \Eprint [0]{\href }%
\providecommand \doibase [0]{https://doi.org/}%
\providecommand \selectlanguage [0]{\@gobble}%
\providecommand \bibinfo  [0]{\@secondoftwo}%
\providecommand \bibfield  [0]{\@secondoftwo}%
\providecommand \translation [1]{[#1]}%
\providecommand \BibitemOpen [0]{}%
\providecommand \bibitemStop [0]{}%
\providecommand \bibitemNoStop [0]{.\EOS\space}%
\providecommand \EOS [0]{\spacefactor3000\relax}%
\providecommand \BibitemShut  [1]{\csname bibitem#1\endcsname}%
\let\auto@bib@innerbib\@empty
\bibitem [{\citenamefont {de~Haas}\ and\ \citenamefont {van
  Alphen}(1930)}]{deHaas1930}%
  \BibitemOpen
  \bibfield  {author} {\bibinfo {author} {\bibfnamefont {W.~J.}\ \bibnamefont
  {de~Haas}}\ and\ \bibinfo {author} {\bibfnamefont {P.~M.}\ \bibnamefont {van
  Alphen}},\ }\bibfield  {title} {\bibinfo {title} {The dependence of the
  susceptibility of diamagnetic metals upon the field},\ }\href
  {https://www.dwc.knaw.nl/DL/publications/PU00015989.pdf} {\bibfield
  {journal} {\bibinfo  {journal} {Proceedings of the Academy of Science of
  Amsterdam}\ }\textbf {\bibinfo {volume} {33}},\ \bibinfo {pages} {1106}
  (\bibinfo {year} {1930})}\BibitemShut {NoStop}%
\bibitem [{\citenamefont {Shubnikov}\ and\ \citenamefont
  {de~Haas}(1930)}]{Shubnikov1930}%
  \BibitemOpen
  \bibfield  {author} {\bibinfo {author} {\bibfnamefont {L.}~\bibnamefont
  {Shubnikov}}\ and\ \bibinfo {author} {\bibfnamefont {W.~J.}\ \bibnamefont
  {de~Haas}},\ }\bibfield  {title} {\bibinfo {title} {Magnetic resistance
  increase in single crystals of bismuth at low temperatures},\ }\href
  {https://www.dwc.knaw.nl/DL/publications/PU00015868.pdf} {\bibfield
  {journal} {\bibinfo  {journal} {Proceedings of the Royal Netherlands Academy
  of Arts and Science}\ }\textbf {\bibinfo {volume} {33}},\ \bibinfo {pages}
  {130} (\bibinfo {year} {1930})}\BibitemShut {NoStop}%
\bibitem [{\citenamefont {Landau}(1930)}]{landau1930diamagnetismus}%
  \BibitemOpen
  \bibfield  {author} {\bibinfo {author} {\bibfnamefont {L.}~\bibnamefont
  {Landau}},\ }\bibfield  {title} {\bibinfo {title} {Diamagnetismus der
  metalle},\ }\href@noop {} {\bibfield  {journal} {\bibinfo  {journal}
  {Zeitschrift f{\"u}r Physik}\ }\textbf {\bibinfo {volume} {64}},\ \bibinfo
  {pages} {629} (\bibinfo {year} {1930})}\BibitemShut {NoStop}%
\bibitem [{\citenamefont {Onsager}(1952)}]{Onsager1952}%
  \BibitemOpen
  \bibfield  {author} {\bibinfo {author} {\bibfnamefont {L.}~\bibnamefont
  {Onsager}},\ }\bibfield  {title} {\bibinfo {title} {Interpretation of the de
  haas-van alphen effect},\ }\href@noop {} {\bibfield  {journal} {\bibinfo
  {journal} {The London, Edinburgh, and Dublin Philosophical Magazine and
  Journal of Science}\ }\textbf {\bibinfo {volume} {43}},\ \bibinfo {pages}
  {1006} (\bibinfo {year} {1952})}\BibitemShut {NoStop}%
\bibitem [{\citenamefont {Lifshitz}\ and\ \citenamefont
  {Kosevich}(1956)}]{Lifshitz1956}%
  \BibitemOpen
  \bibfield  {author} {\bibinfo {author} {\bibfnamefont {I.}~\bibnamefont
  {Lifshitz}}\ and\ \bibinfo {author} {\bibfnamefont {A.}~\bibnamefont
  {Kosevich}},\ }\bibfield  {title} {\bibinfo {title} {Theory of magnetic
  susceptibility in metals at low temperatures},\ }\href@noop {} {\bibfield
  {journal} {\bibinfo  {journal} {Sov. Phys. JETP}\ }\textbf {\bibinfo {volume}
  {2}},\ \bibinfo {pages} {636} (\bibinfo {year} {1956})}\BibitemShut {NoStop}%
\bibitem [{\citenamefont {Shoenberg}(1984)}]{Shoenberg1984}%
  \BibitemOpen
  \bibfield  {author} {\bibinfo {author} {\bibfnamefont {D.}~\bibnamefont
  {Shoenberg}},\ }\href {https://doi.org/10.1017/CBO9780511897870} {\emph
  {\bibinfo {title} {Magnetic Oscillations in Metals}}},\ Cambridge Monographs
  on Physics\ (\bibinfo  {publisher} {Cambridge University Press},\ \bibinfo
  {address} {Cambridge, England},\ \bibinfo {year} {1984})\BibitemShut
  {NoStop}%
\bibitem [{\citenamefont {Tan}\ \emph {et~al.}(2015)\citenamefont {Tan},
  \citenamefont {Hsu}, \citenamefont {Zeng}, \citenamefont {Hatnean},
  \citenamefont {Harrison}, \citenamefont {Zhu}, \citenamefont {Hartstein},
  \citenamefont {Kiourlappou}, \citenamefont {Srivastava}, \citenamefont
  {Johannes} \emph {et~al.}}]{tan2015unconventional}%
  \BibitemOpen
  \bibfield  {author} {\bibinfo {author} {\bibfnamefont {B.}~\bibnamefont
  {Tan}}, \bibinfo {author} {\bibfnamefont {Y.-T.}\ \bibnamefont {Hsu}},
  \bibinfo {author} {\bibfnamefont {B.}~\bibnamefont {Zeng}}, \bibinfo {author}
  {\bibfnamefont {M.~C.}\ \bibnamefont {Hatnean}}, \bibinfo {author}
  {\bibfnamefont {N.}~\bibnamefont {Harrison}}, \bibinfo {author}
  {\bibfnamefont {Z.}~\bibnamefont {Zhu}}, \bibinfo {author} {\bibfnamefont
  {M.}~\bibnamefont {Hartstein}}, \bibinfo {author} {\bibfnamefont
  {M.}~\bibnamefont {Kiourlappou}}, \bibinfo {author} {\bibfnamefont
  {A.}~\bibnamefont {Srivastava}}, \bibinfo {author} {\bibfnamefont
  {M.}~\bibnamefont {Johannes}}, \emph {et~al.},\ }\bibfield  {title} {\bibinfo
  {title} {Unconventional fermi surface in an insulating state},\ }\href@noop
  {} {\bibfield  {journal} {\bibinfo  {journal} {Science}\ }\textbf {\bibinfo
  {volume} {349}},\ \bibinfo {pages} {287} (\bibinfo {year}
  {2015})}\BibitemShut {NoStop}%
\bibitem [{\citenamefont {Hartstein}\ \emph {et~al.}(2018)\citenamefont
  {Hartstein}, \citenamefont {Toews}, \citenamefont {Hsu}, \citenamefont
  {Zeng}, \citenamefont {Chen}, \citenamefont {Hatnean}, \citenamefont {Zhang},
  \citenamefont {Nakamura}, \citenamefont {Padgett}, \citenamefont
  {Rodway-Gant} \emph {et~al.}}]{hartstein2018fermi}%
  \BibitemOpen
  \bibfield  {author} {\bibinfo {author} {\bibfnamefont {M.}~\bibnamefont
  {Hartstein}}, \bibinfo {author} {\bibfnamefont {W.}~\bibnamefont {Toews}},
  \bibinfo {author} {\bibfnamefont {Y.-T.}\ \bibnamefont {Hsu}}, \bibinfo
  {author} {\bibfnamefont {B.}~\bibnamefont {Zeng}}, \bibinfo {author}
  {\bibfnamefont {X.}~\bibnamefont {Chen}}, \bibinfo {author} {\bibfnamefont
  {M.~C.}\ \bibnamefont {Hatnean}}, \bibinfo {author} {\bibfnamefont
  {Q.}~\bibnamefont {Zhang}}, \bibinfo {author} {\bibfnamefont
  {S.}~\bibnamefont {Nakamura}}, \bibinfo {author} {\bibfnamefont
  {A.}~\bibnamefont {Padgett}}, \bibinfo {author} {\bibfnamefont
  {G.}~\bibnamefont {Rodway-Gant}}, \emph {et~al.},\ }\bibfield  {title}
  {\bibinfo {title} {Fermi surface in the absence of a fermi liquid in the
  kondo insulator smb 6},\ }\href@noop {} {\bibfield  {journal} {\bibinfo
  {journal} {Nature Physics}\ }\textbf {\bibinfo {volume} {14}},\ \bibinfo
  {pages} {166} (\bibinfo {year} {2018})}\BibitemShut {NoStop}%
\bibitem [{\citenamefont {Liu}\ \emph {et~al.}(2018)\citenamefont {Liu},
  \citenamefont {Hartstein}, \citenamefont {Wallace}, \citenamefont {Davies},
  \citenamefont {Hatnean}, \citenamefont {Johannes}, \citenamefont
  {Shitsevalova}, \citenamefont {Balakrishnan},\ and\ \citenamefont
  {Sebastian}}]{liu2018fermi}%
  \BibitemOpen
  \bibfield  {author} {\bibinfo {author} {\bibfnamefont {H.}~\bibnamefont
  {Liu}}, \bibinfo {author} {\bibfnamefont {M.}~\bibnamefont {Hartstein}},
  \bibinfo {author} {\bibfnamefont {G.~J.}\ \bibnamefont {Wallace}}, \bibinfo
  {author} {\bibfnamefont {A.~J.}\ \bibnamefont {Davies}}, \bibinfo {author}
  {\bibfnamefont {M.~C.}\ \bibnamefont {Hatnean}}, \bibinfo {author}
  {\bibfnamefont {M.~D.}\ \bibnamefont {Johannes}}, \bibinfo {author}
  {\bibfnamefont {N.}~\bibnamefont {Shitsevalova}}, \bibinfo {author}
  {\bibfnamefont {G.}~\bibnamefont {Balakrishnan}},\ and\ \bibinfo {author}
  {\bibfnamefont {S.~E.}\ \bibnamefont {Sebastian}},\ }\bibfield  {title}
  {\bibinfo {title} {Fermi surfaces in kondo insulators},\ }\href@noop {}
  {\bibfield  {journal} {\bibinfo  {journal} {Journal of Physics: Condensed
  Matter}\ }\textbf {\bibinfo {volume} {30}},\ \bibinfo {pages} {16LT01}
  (\bibinfo {year} {2018})}\BibitemShut {NoStop}%
\bibitem [{\citenamefont {Xiang}\ \emph {et~al.}(2018)\citenamefont {Xiang},
  \citenamefont {Kasahara}, \citenamefont {Asaba}, \citenamefont {Lawson},
  \citenamefont {Tinsman}, \citenamefont {Chen}, \citenamefont {Sugimoto},
  \citenamefont {Kawaguchi}, \citenamefont {Sato}, \citenamefont {Li} \emph
  {et~al.}}]{xiang2018quantum}%
  \BibitemOpen
  \bibfield  {author} {\bibinfo {author} {\bibfnamefont {Z.}~\bibnamefont
  {Xiang}}, \bibinfo {author} {\bibfnamefont {Y.}~\bibnamefont {Kasahara}},
  \bibinfo {author} {\bibfnamefont {T.}~\bibnamefont {Asaba}}, \bibinfo
  {author} {\bibfnamefont {B.}~\bibnamefont {Lawson}}, \bibinfo {author}
  {\bibfnamefont {C.}~\bibnamefont {Tinsman}}, \bibinfo {author} {\bibfnamefont
  {L.}~\bibnamefont {Chen}}, \bibinfo {author} {\bibfnamefont {K.}~\bibnamefont
  {Sugimoto}}, \bibinfo {author} {\bibfnamefont {S.}~\bibnamefont {Kawaguchi}},
  \bibinfo {author} {\bibfnamefont {Y.}~\bibnamefont {Sato}}, \bibinfo {author}
  {\bibfnamefont {G.}~\bibnamefont {Li}}, \emph {et~al.},\ }\bibfield  {title}
  {\bibinfo {title} {Quantum oscillations of electrical resistivity in an
  insulator},\ }\href@noop {} {\bibfield  {journal} {\bibinfo  {journal}
  {Science}\ }\textbf {\bibinfo {volume} {362}},\ \bibinfo {pages} {65}
  (\bibinfo {year} {2018})}\BibitemShut {NoStop}%
\bibitem [{\citenamefont {Hartstein}\ \emph {et~al.}(2020)\citenamefont
  {Hartstein}, \citenamefont {Liu}, \citenamefont {Hsu}, \citenamefont {Tan},
  \citenamefont {Hatnean}, \citenamefont {Balakrishnan},\ and\ \citenamefont
  {Sebastian}}]{hartstein2020intrinsic}%
  \BibitemOpen
  \bibfield  {author} {\bibinfo {author} {\bibfnamefont {M.}~\bibnamefont
  {Hartstein}}, \bibinfo {author} {\bibfnamefont {H.}~\bibnamefont {Liu}},
  \bibinfo {author} {\bibfnamefont {Y.-T.}\ \bibnamefont {Hsu}}, \bibinfo
  {author} {\bibfnamefont {B.~S.}\ \bibnamefont {Tan}}, \bibinfo {author}
  {\bibfnamefont {M.~C.}\ \bibnamefont {Hatnean}}, \bibinfo {author}
  {\bibfnamefont {G.}~\bibnamefont {Balakrishnan}},\ and\ \bibinfo {author}
  {\bibfnamefont {S.~E.}\ \bibnamefont {Sebastian}},\ }\bibfield  {title}
  {\bibinfo {title} {Intrinsic bulk quantum oscillations in a bulk
  unconventional insulator smb6},\ }\href@noop {} {\bibfield  {journal}
  {\bibinfo  {journal} {Iscience}\ }\textbf {\bibinfo {volume} {23}},\ \bibinfo
  {pages} {101632} (\bibinfo {year} {2020})}\BibitemShut {NoStop}%
\bibitem [{\citenamefont {Xiao}\ \emph {et~al.}(2019)\citenamefont {Xiao},
  \citenamefont {Liu}, \citenamefont {Samarth},\ and\ \citenamefont
  {Hu}}]{xiao2019anomalous}%
  \BibitemOpen
  \bibfield  {author} {\bibinfo {author} {\bibfnamefont {D.}~\bibnamefont
  {Xiao}}, \bibinfo {author} {\bibfnamefont {C.-X.}\ \bibnamefont {Liu}},
  \bibinfo {author} {\bibfnamefont {N.}~\bibnamefont {Samarth}},\ and\ \bibinfo
  {author} {\bibfnamefont {L.-H.}\ \bibnamefont {Hu}},\ }\bibfield  {title}
  {\bibinfo {title} {Anomalous quantum oscillations of interacting
  electron-hole gases in inverted type-ii inas/gasb quantum wells},\
  }\href@noop {} {\bibfield  {journal} {\bibinfo  {journal} {Physical review
  letters}\ }\textbf {\bibinfo {volume} {122}},\ \bibinfo {pages} {186802}
  (\bibinfo {year} {2019})}\BibitemShut {NoStop}%
\bibitem [{\citenamefont {Han}\ \emph {et~al.}(2019)\citenamefont {Han},
  \citenamefont {Li}, \citenamefont {Zhang}, \citenamefont {Sullivan},\ and\
  \citenamefont {Du}}]{han2019anomalous}%
  \BibitemOpen
  \bibfield  {author} {\bibinfo {author} {\bibfnamefont {Z.}~\bibnamefont
  {Han}}, \bibinfo {author} {\bibfnamefont {T.}~\bibnamefont {Li}}, \bibinfo
  {author} {\bibfnamefont {L.}~\bibnamefont {Zhang}}, \bibinfo {author}
  {\bibfnamefont {G.}~\bibnamefont {Sullivan}},\ and\ \bibinfo {author}
  {\bibfnamefont {R.-R.}\ \bibnamefont {Du}},\ }\bibfield  {title} {\bibinfo
  {title} {Anomalous conductance oscillations in the hybridization gap of
  inas/gasb quantum wells},\ }\href@noop {} {\bibfield  {journal} {\bibinfo
  {journal} {Physical review letters}\ }\textbf {\bibinfo {volume} {123}},\
  \bibinfo {pages} {126803} (\bibinfo {year} {2019})}\BibitemShut {NoStop}%
\bibitem [{\citenamefont {Wang}\ \emph {et~al.}(2021)\citenamefont {Wang},
  \citenamefont {Yu}, \citenamefont {Jia}, \citenamefont {Onyszczak},
  \citenamefont {Cevallos}, \citenamefont {Lei}, \citenamefont {Klemenz},
  \citenamefont {Watanabe}, \citenamefont {Taniguchi}, \citenamefont {Cava}
  \emph {et~al.}}]{wang2021landau}%
  \BibitemOpen
  \bibfield  {author} {\bibinfo {author} {\bibfnamefont {P.}~\bibnamefont
  {Wang}}, \bibinfo {author} {\bibfnamefont {G.}~\bibnamefont {Yu}}, \bibinfo
  {author} {\bibfnamefont {Y.}~\bibnamefont {Jia}}, \bibinfo {author}
  {\bibfnamefont {M.}~\bibnamefont {Onyszczak}}, \bibinfo {author}
  {\bibfnamefont {F.~A.}\ \bibnamefont {Cevallos}}, \bibinfo {author}
  {\bibfnamefont {S.}~\bibnamefont {Lei}}, \bibinfo {author} {\bibfnamefont
  {S.}~\bibnamefont {Klemenz}}, \bibinfo {author} {\bibfnamefont
  {K.}~\bibnamefont {Watanabe}}, \bibinfo {author} {\bibfnamefont
  {T.}~\bibnamefont {Taniguchi}}, \bibinfo {author} {\bibfnamefont {R.~J.}\
  \bibnamefont {Cava}}, \emph {et~al.},\ }\bibfield  {title} {\bibinfo {title}
  {Landau quantization and highly mobile fermions in an insulator},\
  }\href@noop {} {\bibfield  {journal} {\bibinfo  {journal} {Nature}\ }\textbf
  {\bibinfo {volume} {589}},\ \bibinfo {pages} {225} (\bibinfo {year}
  {2021})}\BibitemShut {NoStop}%
\bibitem [{\citenamefont {Leeb}\ \emph {et~al.}(2021)\citenamefont {Leeb},
  \citenamefont {Polyudov}, \citenamefont {Mashhadi}, \citenamefont {Biswas},
  \citenamefont {Valent{\'\i}}, \citenamefont {Burghard},\ and\ \citenamefont
  {Knolle}}]{leeb2021anomalous}%
  \BibitemOpen
  \bibfield  {author} {\bibinfo {author} {\bibfnamefont {V.}~\bibnamefont
  {Leeb}}, \bibinfo {author} {\bibfnamefont {K.}~\bibnamefont {Polyudov}},
  \bibinfo {author} {\bibfnamefont {S.}~\bibnamefont {Mashhadi}}, \bibinfo
  {author} {\bibfnamefont {S.}~\bibnamefont {Biswas}}, \bibinfo {author}
  {\bibfnamefont {R.}~\bibnamefont {Valent{\'\i}}}, \bibinfo {author}
  {\bibfnamefont {M.}~\bibnamefont {Burghard}},\ and\ \bibinfo {author}
  {\bibfnamefont {J.}~\bibnamefont {Knolle}},\ }\bibfield  {title} {\bibinfo
  {title} {Anomalous quantum oscillations in a heterostructure of graphene on a
  proximate quantum spin liquid},\ }\href@noop {} {\bibfield  {journal}
  {\bibinfo  {journal} {Physical Review Letters}\ }\textbf {\bibinfo {volume}
  {126}},\ \bibinfo {pages} {097201} (\bibinfo {year} {2021})}\BibitemShut
  {NoStop}%
\bibitem [{\citenamefont {Knolle}\ and\ \citenamefont
  {Cooper}(2017)}]{knolle2017anomalous}%
  \BibitemOpen
  \bibfield  {author} {\bibinfo {author} {\bibfnamefont {J.}~\bibnamefont
  {Knolle}}\ and\ \bibinfo {author} {\bibfnamefont {N.~R.}\ \bibnamefont
  {Cooper}},\ }\bibfield  {title} {\bibinfo {title} {Anomalous de haas--van
  alphen effect in inas/gasb quantum wells},\ }\href@noop {} {\bibfield
  {journal} {\bibinfo  {journal} {Physical Review Letters}\ }\textbf {\bibinfo
  {volume} {118}},\ \bibinfo {pages} {176801} (\bibinfo {year}
  {2017})}\BibitemShut {NoStop}%
\bibitem [{\citenamefont {Shen}\ and\ \citenamefont
  {Fu}(2018)}]{shen2018quantum}%
  \BibitemOpen
  \bibfield  {author} {\bibinfo {author} {\bibfnamefont {H.}~\bibnamefont
  {Shen}}\ and\ \bibinfo {author} {\bibfnamefont {L.}~\bibnamefont {Fu}},\
  }\bibfield  {title} {\bibinfo {title} {Quantum oscillation from in-gap states
  and a non-hermitian landau level problem},\ }\href@noop {} {\bibfield
  {journal} {\bibinfo  {journal} {Physical review letters}\ }\textbf {\bibinfo
  {volume} {121}},\ \bibinfo {pages} {026403} (\bibinfo {year}
  {2018})}\BibitemShut {NoStop}%
\bibitem [{\citenamefont {Erten}\ \emph {et~al.}(2016)\citenamefont {Erten},
  \citenamefont {Ghaemi},\ and\ \citenamefont {Coleman}}]{erten2016kondo}%
  \BibitemOpen
  \bibfield  {author} {\bibinfo {author} {\bibfnamefont {O.}~\bibnamefont
  {Erten}}, \bibinfo {author} {\bibfnamefont {P.}~\bibnamefont {Ghaemi}},\ and\
  \bibinfo {author} {\bibfnamefont {P.}~\bibnamefont {Coleman}},\ }\bibfield
  {title} {\bibinfo {title} {Kondo breakdown and quantum oscillations in smb
  6},\ }\href@noop {} {\bibfield  {journal} {\bibinfo  {journal} {Physical
  review letters}\ }\textbf {\bibinfo {volume} {116}},\ \bibinfo {pages}
  {046403} (\bibinfo {year} {2016})}\BibitemShut {NoStop}%
\bibitem [{\citenamefont {Zhang}\ \emph {et~al.}(2016)\citenamefont {Zhang},
  \citenamefont {Song},\ and\ \citenamefont {Wang}}]{zhang2016quantum}%
  \BibitemOpen
  \bibfield  {author} {\bibinfo {author} {\bibfnamefont {L.}~\bibnamefont
  {Zhang}}, \bibinfo {author} {\bibfnamefont {X.-Y.}\ \bibnamefont {Song}},\
  and\ \bibinfo {author} {\bibfnamefont {F.}~\bibnamefont {Wang}},\ }\bibfield
  {title} {\bibinfo {title} {Quantum oscillation in narrow-gap topological
  insulators},\ }\href@noop {} {\bibfield  {journal} {\bibinfo  {journal}
  {Physical review letters}\ }\textbf {\bibinfo {volume} {116}},\ \bibinfo
  {pages} {046404} (\bibinfo {year} {2016})}\BibitemShut {NoStop}%
\bibitem [{\citenamefont {Sodemann}\ \emph {et~al.}(2018)\citenamefont
  {Sodemann}, \citenamefont {Chowdhury},\ and\ \citenamefont
  {Senthil}}]{sodemann2018quantum}%
  \BibitemOpen
  \bibfield  {author} {\bibinfo {author} {\bibfnamefont {I.}~\bibnamefont
  {Sodemann}}, \bibinfo {author} {\bibfnamefont {D.}~\bibnamefont
  {Chowdhury}},\ and\ \bibinfo {author} {\bibfnamefont {T.}~\bibnamefont
  {Senthil}},\ }\bibfield  {title} {\bibinfo {title} {Quantum oscillations in
  insulators with neutral fermi surfaces},\ }\href@noop {} {\bibfield
  {journal} {\bibinfo  {journal} {Physical Review B}\ }\textbf {\bibinfo
  {volume} {97}},\ \bibinfo {pages} {045152} (\bibinfo {year}
  {2018})}\BibitemShut {NoStop}%
\bibitem [{\citenamefont {Lee}(2021)}]{lee2021quantum}%
  \BibitemOpen
  \bibfield  {author} {\bibinfo {author} {\bibfnamefont {P.~A.}\ \bibnamefont
  {Lee}},\ }\bibfield  {title} {\bibinfo {title} {Quantum oscillations in the
  activated conductivity in excitonic insulators: Possible application to
  monolayer wte 2},\ }\href@noop {} {\bibfield  {journal} {\bibinfo  {journal}
  {Physical Review B}\ }\textbf {\bibinfo {volume} {103}},\ \bibinfo {pages}
  {L041101} (\bibinfo {year} {2021})}\BibitemShut {NoStop}%
\bibitem [{\citenamefont {He}\ and\ \citenamefont {Lee}(2021)}]{he2021quantum}%
  \BibitemOpen
  \bibfield  {author} {\bibinfo {author} {\bibfnamefont {W.-Y.}\ \bibnamefont
  {He}}\ and\ \bibinfo {author} {\bibfnamefont {P.~A.}\ \bibnamefont {Lee}},\
  }\bibfield  {title} {\bibinfo {title} {Quantum oscillation of thermally
  activated conductivity in a monolayer wte $ \_2 $-like excitonic insulator},\
  }\href@noop {} {\bibfield  {journal} {\bibinfo  {journal} {arXiv preprint
  arXiv:2105.02411}\ } (\bibinfo {year} {2021})}\BibitemShut {NoStop}%
\bibitem [{\citenamefont {Cohen}\ and\ \citenamefont
  {Falicov}(1961)}]{Cohen1961}%
  \BibitemOpen
  \bibfield  {author} {\bibinfo {author} {\bibfnamefont {M.~H.}\ \bibnamefont
  {Cohen}}\ and\ \bibinfo {author} {\bibfnamefont {L.~M.}\ \bibnamefont
  {Falicov}},\ }\bibfield  {title} {\bibinfo {title} {Magnetic breakdown in
  crystals},\ }\href {https://doi.org/10.1103/PhysRevLett.7.231} {\bibfield
  {journal} {\bibinfo  {journal} {Phys. Rev. Lett.}\ }\textbf {\bibinfo
  {volume} {7}},\ \bibinfo {pages} {231} (\bibinfo {year} {1961})}\BibitemShut
  {NoStop}%
\bibitem [{\citenamefont {Blount}(1962)}]{Blount1962}%
  \BibitemOpen
  \bibfield  {author} {\bibinfo {author} {\bibfnamefont {E.~I.}\ \bibnamefont
  {Blount}},\ }\bibfield  {title} {\bibinfo {title} {Bloch electrons in a
  magnetic field},\ }\href {https://doi.org/10.1103/PhysRev.126.1636}
  {\bibfield  {journal} {\bibinfo  {journal} {Phys. Rev.}\ }\textbf {\bibinfo
  {volume} {126}},\ \bibinfo {pages} {1636} (\bibinfo {year}
  {1962})}\BibitemShut {NoStop}%
\bibitem [{\citenamefont {Slutskin}(1968)}]{slutskin1968dynamics}%
  \BibitemOpen
  \bibfield  {author} {\bibinfo {author} {\bibfnamefont {A.}~\bibnamefont
  {Slutskin}},\ }\bibfield  {title} {\bibinfo {title} {Dynamics of conduction
  electrons under magnetic breakdown conditions},\ }\href@noop {} {\bibfield
  {journal} {\bibinfo  {journal} {Sov. Phys. JETP}\ }\textbf {\bibinfo {volume}
  {26}},\ \bibinfo {pages} {474} (\bibinfo {year} {1968})}\BibitemShut
  {NoStop}%
\bibitem [{\citenamefont {Alexandradinata}\ and\ \citenamefont
  {Glazman}(2018)}]{alexandradinata2018semiclassical}%
  \BibitemOpen
  \bibfield  {author} {\bibinfo {author} {\bibfnamefont {A.}~\bibnamefont
  {Alexandradinata}}\ and\ \bibinfo {author} {\bibfnamefont {L.}~\bibnamefont
  {Glazman}},\ }\bibfield  {title} {\bibinfo {title} {Semiclassical theory of
  landau levels and magnetic breakdown in topological metals},\ }\href@noop {}
  {\bibfield  {journal} {\bibinfo  {journal} {Physical Review B}\ }\textbf
  {\bibinfo {volume} {97}},\ \bibinfo {pages} {144422} (\bibinfo {year}
  {2018})}\BibitemShut {NoStop}%
\bibitem [{\citenamefont {O'Brien}\ \emph {et~al.}(2016)\citenamefont
  {O'Brien}, \citenamefont {Diez},\ and\ \citenamefont
  {Beenakker}}]{OBrien2016}%
  \BibitemOpen
  \bibfield  {author} {\bibinfo {author} {\bibfnamefont {T.~E.}\ \bibnamefont
  {O'Brien}}, \bibinfo {author} {\bibfnamefont {M.}~\bibnamefont {Diez}},\ and\
  \bibinfo {author} {\bibfnamefont {C.~W.~J.}\ \bibnamefont {Beenakker}},\
  }\bibfield  {title} {\bibinfo {title} {Magnetic breakdown and klein tunneling
  in a type-ii weyl semimetal},\ }\href
  {https://doi.org/10.1103/PhysRevLett.116.236401} {\bibfield  {journal}
  {\bibinfo  {journal} {Phys. Rev. Lett.}\ }\textbf {\bibinfo {volume} {116}},\
  \bibinfo {pages} {236401} (\bibinfo {year} {2016})}\BibitemShut {NoStop}%
\bibitem [{\citenamefont {Alexandradinata}\ and\ \citenamefont
  {Glazman}(2017)}]{alexandradinata2017geometric}%
  \BibitemOpen
  \bibfield  {author} {\bibinfo {author} {\bibfnamefont {A.}~\bibnamefont
  {Alexandradinata}}\ and\ \bibinfo {author} {\bibfnamefont {L.}~\bibnamefont
  {Glazman}},\ }\bibfield  {title} {\bibinfo {title} {Geometric phase and
  orbital moment in quantization rules for magnetic breakdown},\ }\href@noop {}
  {\bibfield  {journal} {\bibinfo  {journal} {Physical review letters}\
  }\textbf {\bibinfo {volume} {119}},\ \bibinfo {pages} {256601} (\bibinfo
  {year} {2017})}\BibitemShut {NoStop}%
\bibitem [{\citenamefont {Van~Delft}\ \emph {et~al.}(2018)\citenamefont
  {Van~Delft}, \citenamefont {Pezzini}, \citenamefont {Khouri}, \citenamefont
  {M{\"u}ller}, \citenamefont {Breitkreiz}, \citenamefont {Schoop},
  \citenamefont {Carrington}, \citenamefont {Hussey},\ and\ \citenamefont
  {Wiedmann}}]{van2018electron}%
  \BibitemOpen
  \bibfield  {author} {\bibinfo {author} {\bibfnamefont {M.}~\bibnamefont
  {Van~Delft}}, \bibinfo {author} {\bibfnamefont {S.}~\bibnamefont {Pezzini}},
  \bibinfo {author} {\bibfnamefont {T.}~\bibnamefont {Khouri}}, \bibinfo
  {author} {\bibfnamefont {C.}~\bibnamefont {M{\"u}ller}}, \bibinfo {author}
  {\bibfnamefont {M.}~\bibnamefont {Breitkreiz}}, \bibinfo {author}
  {\bibfnamefont {L.~M.}\ \bibnamefont {Schoop}}, \bibinfo {author}
  {\bibfnamefont {A.}~\bibnamefont {Carrington}}, \bibinfo {author}
  {\bibfnamefont {N.}~\bibnamefont {Hussey}},\ and\ \bibinfo {author}
  {\bibfnamefont {S.}~\bibnamefont {Wiedmann}},\ }\bibfield  {title} {\bibinfo
  {title} {Electron-hole tunneling revealed by quantum oscillations in the
  nodal-line semimetal hfsis},\ }\href@noop {} {\bibfield  {journal} {\bibinfo
  {journal} {Physical review letters}\ }\textbf {\bibinfo {volume} {121}},\
  \bibinfo {pages} {256602} (\bibinfo {year} {2018})}\BibitemShut {NoStop}%
\bibitem [{\citenamefont {Polyanovsky}(1988)}]{Polyanovsky1988}%
  \BibitemOpen
  \bibfield  {author} {\bibinfo {author} {\bibfnamefont {V.}~\bibnamefont
  {Polyanovsky}},\ }\bibfield  {title} {\bibinfo {title} {Magnetointersubband
  oscillations of conductivity in a two-dimensional electronic system},\
  }\href@noop {} {\bibfield  {journal} {\bibinfo  {journal} {Fiz. Tekh.
  Poluprovodn.}\ }\textbf {\bibinfo {volume} {22}},\ \bibinfo {pages} {1408}
  (\bibinfo {year} {1988})}\BibitemShut {NoStop}%
\bibitem [{\citenamefont {Raikh}\ and\ \citenamefont
  {Shahbazyan}(1994)}]{Raikh1994}%
  \BibitemOpen
  \bibfield  {author} {\bibinfo {author} {\bibfnamefont {M.~E.}\ \bibnamefont
  {Raikh}}\ and\ \bibinfo {author} {\bibfnamefont {T.~V.}\ \bibnamefont
  {Shahbazyan}},\ }\bibfield  {title} {\bibinfo {title} {Magnetointersubband
  oscillations of conductivity in a two-dimensional electronic system},\ }\href
  {https://doi.org/10.1103/PhysRevB.49.5531} {\bibfield  {journal} {\bibinfo
  {journal} {Phys. Rev. B}\ }\textbf {\bibinfo {volume} {49}},\ \bibinfo
  {pages} {5531} (\bibinfo {year} {1994})}\BibitemShut {NoStop}%
\bibitem [{\citenamefont {Averkiev}\ \emph {et~al.}(2001)\citenamefont
  {Averkiev}, \citenamefont {Golub}, \citenamefont {Tarasenko},\ and\
  \citenamefont {Willander}}]{Averkiev2001}%
  \BibitemOpen
  \bibfield  {author} {\bibinfo {author} {\bibfnamefont {N.~S.}\ \bibnamefont
  {Averkiev}}, \bibinfo {author} {\bibfnamefont {L.~E.}\ \bibnamefont {Golub}},
  \bibinfo {author} {\bibfnamefont {S.~A.}\ \bibnamefont {Tarasenko}},\ and\
  \bibinfo {author} {\bibfnamefont {M.}~\bibnamefont {Willander}},\ }\bibfield
  {title} {\bibinfo {title} {Theory of magneto-oscillation effects in
  quasi-two-dimensional semiconductor structures},\ }\href
  {https://doi.org/10.1088/0953-8984/13/11/309} {\bibfield  {journal} {\bibinfo
   {journal} {Journal of Physics: Condensed Matter}\ }\textbf {\bibinfo
  {volume} {13}},\ \bibinfo {pages} {2517} (\bibinfo {year}
  {2001})}\BibitemShut {NoStop}%
\bibitem [{\citenamefont {Coleridge}(1990)}]{Coleridge1990}%
  \BibitemOpen
  \bibfield  {author} {\bibinfo {author} {\bibfnamefont {P.~T.}\ \bibnamefont
  {Coleridge}},\ }\bibfield  {title} {\bibinfo {title} {Inter-subband
  scattering in a 2d electron gas},\ }\href
  {https://doi.org/10.1088/0268-1242/5/9/006} {\bibfield  {journal} {\bibinfo
  {journal} {Semiconductor Science and Technology}\ }\textbf {\bibinfo {volume}
  {5}},\ \bibinfo {pages} {961} (\bibinfo {year} {1990})}\BibitemShut {NoStop}%
\bibitem [{\citenamefont {Leadley}\ \emph {et~al.}(1992)\citenamefont
  {Leadley}, \citenamefont {Fletcher}, \citenamefont {Nicholas}, \citenamefont
  {Tao}, \citenamefont {Foxon},\ and\ \citenamefont {Harris}}]{Leadley1992}%
  \BibitemOpen
  \bibfield  {author} {\bibinfo {author} {\bibfnamefont {D.~R.}\ \bibnamefont
  {Leadley}}, \bibinfo {author} {\bibfnamefont {R.}~\bibnamefont {Fletcher}},
  \bibinfo {author} {\bibfnamefont {R.~J.}\ \bibnamefont {Nicholas}}, \bibinfo
  {author} {\bibfnamefont {F.}~\bibnamefont {Tao}}, \bibinfo {author}
  {\bibfnamefont {C.~T.}\ \bibnamefont {Foxon}},\ and\ \bibinfo {author}
  {\bibfnamefont {J.~J.}\ \bibnamefont {Harris}},\ }\bibfield  {title}
  {\bibinfo {title} {Intersubband resonant scattering in
  {GaAs-${\mathrm{Ga}}_{1\mathrm{\ensuremath{-}}\mathit{x}}$
  ${\mathrm{Al}}_{\mathit{x}}$As} heterojunctions},\ }\href
  {https://doi.org/10.1103/PhysRevB.46.12439} {\bibfield  {journal} {\bibinfo
  {journal} {Phys. Rev. B}\ }\textbf {\bibinfo {volume} {46}},\ \bibinfo
  {pages} {12439} (\bibinfo {year} {1992})}\BibitemShut {NoStop}%
\bibitem [{\citenamefont {Goran}\ \emph {et~al.}(2009)\citenamefont {Goran},
  \citenamefont {Bykov}, \citenamefont {Toropov},\ and\ \citenamefont
  {Vitkalov}}]{Goran2009}%
  \BibitemOpen
  \bibfield  {author} {\bibinfo {author} {\bibfnamefont {A.~V.}\ \bibnamefont
  {Goran}}, \bibinfo {author} {\bibfnamefont {A.~A.}\ \bibnamefont {Bykov}},
  \bibinfo {author} {\bibfnamefont {A.~I.}\ \bibnamefont {Toropov}},\ and\
  \bibinfo {author} {\bibfnamefont {S.~A.}\ \bibnamefont {Vitkalov}},\
  }\bibfield  {title} {\bibinfo {title} {Effect of electron-electron scattering
  on magnetointersubband resistance oscillations of two-dimensional electrons
  in {GaAs} quantum wells},\ }\href
  {https://doi.org/10.1103/PhysRevB.80.193305} {\bibfield  {journal} {\bibinfo
  {journal} {Phys. Rev. B}\ }\textbf {\bibinfo {volume} {80}},\ \bibinfo
  {pages} {193305} (\bibinfo {year} {2009})}\BibitemShut {NoStop}%
\bibitem [{\citenamefont {Polyanovsky}(1993)}]{Polyanosky1993}%
  \BibitemOpen
  \bibfield  {author} {\bibinfo {author} {\bibfnamefont {V.}~\bibnamefont
  {Polyanovsky}},\ }\bibfield  {title} {\bibinfo {title} {High-temperature
  quantum oscillations of the magnetoresistance in layered systems},\ }\href
  {https://doi.org/10.1103/PhysRevB.47.1985} {\bibfield  {journal} {\bibinfo
  {journal} {Phys. Rev. B}\ }\textbf {\bibinfo {volume} {47}},\ \bibinfo
  {pages} {1985} (\bibinfo {year} {1993})}\BibitemShut {NoStop}%
\bibitem [{\citenamefont {Grigoriev}(2003)}]{Grigoriev2003}%
  \BibitemOpen
  \bibfield  {author} {\bibinfo {author} {\bibfnamefont {P.~D.}\ \bibnamefont
  {Grigoriev}},\ }\bibfield  {title} {\bibinfo {title} {Theory of the
  shubnikov--de haas effect in quasi-two-dimensional metals},\ }\href
  {https://doi.org/10.1103/PhysRevB.67.144401} {\bibfield  {journal} {\bibinfo
  {journal} {Phys. Rev. B}\ }\textbf {\bibinfo {volume} {67}},\ \bibinfo
  {pages} {144401} (\bibinfo {year} {2003})}\BibitemShut {NoStop}%
\bibitem [{\citenamefont {Thomas}\ \emph {et~al.}(2008)\citenamefont {Thomas},
  \citenamefont {Kabanov},\ and\ \citenamefont {Alexandrov}}]{Thomas2008}%
  \BibitemOpen
  \bibfield  {author} {\bibinfo {author} {\bibfnamefont {I.~O.}\ \bibnamefont
  {Thomas}}, \bibinfo {author} {\bibfnamefont {V.~V.}\ \bibnamefont
  {Kabanov}},\ and\ \bibinfo {author} {\bibfnamefont {A.~S.}\ \bibnamefont
  {Alexandrov}},\ }\bibfield  {title} {\bibinfo {title} {Shubnikov--de haas
  effect in multiband quasi-two-dimensional metals},\ }\href
  {https://doi.org/10.1103/PhysRevB.77.075434} {\bibfield  {journal} {\bibinfo
  {journal} {Phys. Rev. B}\ }\textbf {\bibinfo {volume} {77}},\ \bibinfo
  {pages} {075434} (\bibinfo {year} {2008})}\BibitemShut {NoStop}%
\bibitem [{\citenamefont {Mogilyuk}\ and\ \citenamefont
  {Grigoriev}(2018)}]{Mogilyuk2018}%
  \BibitemOpen
  \bibfield  {author} {\bibinfo {author} {\bibfnamefont {T.~I.}\ \bibnamefont
  {Mogilyuk}}\ and\ \bibinfo {author} {\bibfnamefont {P.~D.}\ \bibnamefont
  {Grigoriev}},\ }\bibfield  {title} {\bibinfo {title} {Magnetic oscillations
  of in-plane conductivity in quasi-two-dimensional metals},\ }\href
  {https://doi.org/10.1103/PhysRevB.98.045118} {\bibfield  {journal} {\bibinfo
  {journal} {Phys. Rev. B}\ }\textbf {\bibinfo {volume} {98}},\ \bibinfo
  {pages} {045118} (\bibinfo {year} {2018})}\BibitemShut {NoStop}%
\bibitem [{\citenamefont {Manes}(2012)}]{manes2012existence}%
  \BibitemOpen
  \bibfield  {author} {\bibinfo {author} {\bibfnamefont {J.~L.}\ \bibnamefont
  {Manes}},\ }\bibfield  {title} {\bibinfo {title} {Existence of bulk chiral
  fermions and crystal symmetry},\ }\href@noop {} {\bibfield  {journal}
  {\bibinfo  {journal} {Physical Review B}\ }\textbf {\bibinfo {volume} {85}},\
  \bibinfo {pages} {155118} (\bibinfo {year} {2012})}\BibitemShut {NoStop}%
\bibitem [{\citenamefont {Fang}\ \emph {et~al.}(2012)\citenamefont {Fang},
  \citenamefont {Gilbert}, \citenamefont {Dai},\ and\ \citenamefont
  {Bernevig}}]{Bernevig2012}%
  \BibitemOpen
  \bibfield  {author} {\bibinfo {author} {\bibfnamefont {C.}~\bibnamefont
  {Fang}}, \bibinfo {author} {\bibfnamefont {M.~J.}\ \bibnamefont {Gilbert}},
  \bibinfo {author} {\bibfnamefont {X.}~\bibnamefont {Dai}},\ and\ \bibinfo
  {author} {\bibfnamefont {B.~A.}\ \bibnamefont {Bernevig}},\ }\bibfield
  {title} {\bibinfo {title} {Multi-weyl topological semimetals stabilized by
  point group symmetry},\ }\href
  {https://doi.org/10.1103/PhysRevLett.108.266802} {\bibfield  {journal}
  {\bibinfo  {journal} {Phys. Rev. Lett.}\ }\textbf {\bibinfo {volume} {108}},\
  \bibinfo {pages} {266802} (\bibinfo {year} {2012})}\BibitemShut {NoStop}%
\bibitem [{\citenamefont {Bradlyn}\ \emph {et~al.}(2016)\citenamefont
  {Bradlyn}, \citenamefont {Cano}, \citenamefont {Wang}, \citenamefont
  {Vergniory}, \citenamefont {Felser}, \citenamefont {Cava},\ and\
  \citenamefont {Bernevig}}]{bradlyn2016beyond}%
  \BibitemOpen
  \bibfield  {author} {\bibinfo {author} {\bibfnamefont {B.}~\bibnamefont
  {Bradlyn}}, \bibinfo {author} {\bibfnamefont {J.}~\bibnamefont {Cano}},
  \bibinfo {author} {\bibfnamefont {Z.}~\bibnamefont {Wang}}, \bibinfo {author}
  {\bibfnamefont {M.}~\bibnamefont {Vergniory}}, \bibinfo {author}
  {\bibfnamefont {C.}~\bibnamefont {Felser}}, \bibinfo {author} {\bibfnamefont
  {R.~J.}\ \bibnamefont {Cava}},\ and\ \bibinfo {author} {\bibfnamefont
  {B.~A.}\ \bibnamefont {Bernevig}},\ }\bibfield  {title} {\bibinfo {title}
  {Beyond dirac and weyl fermions: Unconventional quasiparticles in
  conventional crystals},\ }\href@noop {} {\bibfield  {journal} {\bibinfo
  {journal} {Science}\ }\textbf {\bibinfo {volume} {353}} (\bibinfo {year}
  {2016})}\BibitemShut {NoStop}%
\bibitem [{\citenamefont {Tang}\ \emph {et~al.}(2017)\citenamefont {Tang},
  \citenamefont {Zhou},\ and\ \citenamefont {Zhang}}]{tang2017multiple}%
  \BibitemOpen
  \bibfield  {author} {\bibinfo {author} {\bibfnamefont {P.}~\bibnamefont
  {Tang}}, \bibinfo {author} {\bibfnamefont {Q.}~\bibnamefont {Zhou}},\ and\
  \bibinfo {author} {\bibfnamefont {S.-C.}\ \bibnamefont {Zhang}},\ }\bibfield
  {title} {\bibinfo {title} {Multiple types of topological fermions in
  transition metal silicides},\ }\href@noop {} {\bibfield  {journal} {\bibinfo
  {journal} {Physical review letters}\ }\textbf {\bibinfo {volume} {119}},\
  \bibinfo {pages} {206402} (\bibinfo {year} {2017})}\BibitemShut {NoStop}%
\bibitem [{\citenamefont {Huber}\ \emph {et~al.}(2023)\citenamefont {Huber},
  \citenamefont {Leeb}, \citenamefont {Bauer}, \citenamefont {Benka},
  \citenamefont {Knolle}, \citenamefont {Pfleiderer},\ and\ \citenamefont
  {Wilde}}]{Huber2023}%
  \BibitemOpen
  \bibfield  {author} {\bibinfo {author} {\bibfnamefont {N.}~\bibnamefont
  {Huber}}, \bibinfo {author} {\bibfnamefont {V.}~\bibnamefont {Leeb}},
  \bibinfo {author} {\bibfnamefont {A.}~\bibnamefont {Bauer}}, \bibinfo
  {author} {\bibfnamefont {G.}~\bibnamefont {Benka}}, \bibinfo {author}
  {\bibfnamefont {J.}~\bibnamefont {Knolle}}, \bibinfo {author} {\bibfnamefont
  {C.}~\bibnamefont {Pfleiderer}},\ and\ \bibinfo {author} {\bibfnamefont
  {M.~A.}\ \bibnamefont {Wilde}},\ }\href@noop {} {\bibinfo {title} {Quantum
  oscillations of the quasiparticle lifetime in a metal}} (\bibinfo {year}
  {2023}),\ \Eprint {https://arxiv.org/abs/2306.09420} {arXiv:2306.09420
  [cond-mat.str-el]} \BibitemShut {NoStop}%
\bibitem [{\citenamefont {Peres}\ \emph {et~al.}(2006)\citenamefont {Peres},
  \citenamefont {Guinea},\ and\ \citenamefont {Castro~Neto}}]{Peres2006}%
  \BibitemOpen
  \bibfield  {author} {\bibinfo {author} {\bibfnamefont {N.~M.~R.}\
  \bibnamefont {Peres}}, \bibinfo {author} {\bibfnamefont {F.}~\bibnamefont
  {Guinea}},\ and\ \bibinfo {author} {\bibfnamefont {A.~H.}\ \bibnamefont
  {Castro~Neto}},\ }\bibfield  {title} {\bibinfo {title} {Electronic properties
  of disordered two-dimensional carbon},\ }\href
  {https://doi.org/10.1103/PhysRevB.73.125411} {\bibfield  {journal} {\bibinfo
  {journal} {Phys. Rev. B}\ }\textbf {\bibinfo {volume} {73}},\ \bibinfo
  {pages} {125411} (\bibinfo {year} {2006})}\BibitemShut {NoStop}%
\bibitem [{\citenamefont {Leadley}\ \emph {et~al.}(1989)\citenamefont
  {Leadley}, \citenamefont {Nicholas}, \citenamefont {Harris},\ and\
  \citenamefont {Foxon}}]{Leadley1989}%
  \BibitemOpen
  \bibfield  {author} {\bibinfo {author} {\bibfnamefont {D.~R.}\ \bibnamefont
  {Leadley}}, \bibinfo {author} {\bibfnamefont {R.~J.}\ \bibnamefont
  {Nicholas}}, \bibinfo {author} {\bibfnamefont {J.~J.}\ \bibnamefont
  {Harris}},\ and\ \bibinfo {author} {\bibfnamefont {C.~T.}\ \bibnamefont
  {Foxon}},\ }\bibfield  {title} {\bibinfo {title} {Influence of acoustic
  phonons on inter-subband scattering in {GaAs}-{GaAlAs} heterojunctions},\
  }\href {https://doi.org/10.1088/0268-1242/4/10/010} {\bibfield  {journal}
  {\bibinfo  {journal} {Semiconductor Science and Technology}\ }\textbf
  {\bibinfo {volume} {4}},\ \bibinfo {pages} {885} (\bibinfo {year}
  {1989})}\BibitemShut {NoStop}%
\bibitem [{\citenamefont {Sander}\ \emph {et~al.}(1996)\citenamefont {Sander},
  \citenamefont {Holmes}, \citenamefont {Harris}, \citenamefont {Maude},\ and\
  \citenamefont {Portal}}]{Sander1996}%
  \BibitemOpen
  \bibfield  {author} {\bibinfo {author} {\bibfnamefont {T.}~\bibnamefont
  {Sander}}, \bibinfo {author} {\bibfnamefont {S.}~\bibnamefont {Holmes}},
  \bibinfo {author} {\bibfnamefont {J.}~\bibnamefont {Harris}}, \bibinfo
  {author} {\bibfnamefont {D.}~\bibnamefont {Maude}},\ and\ \bibinfo {author}
  {\bibfnamefont {J.}~\bibnamefont {Portal}},\ }\bibfield  {title} {\bibinfo
  {title} {Magnetoresistance oscillations due to intersubband scattering in a
  two-dimensional electron system},\ }\href
  {https://doi.org/https://doi.org/10.1016/0039-6028(96)00470-0} {\bibfield
  {journal} {\bibinfo  {journal} {Surface Science}\ }\textbf {\bibinfo {volume}
  {361-362}},\ \bibinfo {pages} {564} (\bibinfo {year} {1996})}\BibitemShut
  {NoStop}%
\bibitem [{\citenamefont {Minkov}\ \emph {et~al.}(2020)\citenamefont {Minkov},
  \citenamefont {Rut}, \citenamefont {Sherstobitov}, \citenamefont {Dvoretski},
  \citenamefont {Mikhailov}, \citenamefont {Solov'ev}, \citenamefont {Chernov},
  \citenamefont {Ivanov},\ and\ \citenamefont {Germanenko}}]{Minkov2020}%
  \BibitemOpen
  \bibfield  {author} {\bibinfo {author} {\bibfnamefont {G.~M.}\ \bibnamefont
  {Minkov}}, \bibinfo {author} {\bibfnamefont {O.~E.}\ \bibnamefont {Rut}},
  \bibinfo {author} {\bibfnamefont {A.~A.}\ \bibnamefont {Sherstobitov}},
  \bibinfo {author} {\bibfnamefont {S.~A.}\ \bibnamefont {Dvoretski}}, \bibinfo
  {author} {\bibfnamefont {N.~N.}\ \bibnamefont {Mikhailov}}, \bibinfo {author}
  {\bibfnamefont {V.~A.}\ \bibnamefont {Solov'ev}}, \bibinfo {author}
  {\bibfnamefont {M.~Y.}\ \bibnamefont {Chernov}}, \bibinfo {author}
  {\bibfnamefont {S.~V.}\ \bibnamefont {Ivanov}},\ and\ \bibinfo {author}
  {\bibfnamefont {A.~V.}\ \bibnamefont {Germanenko}},\ }\bibfield  {title}
  {\bibinfo {title} {Magneto-intersubband oscillations in two-dimensional
  systems with an energy spectrum split due to spin-orbit interaction},\ }\href
  {https://doi.org/10.1103/PhysRevB.101.245303} {\bibfield  {journal} {\bibinfo
   {journal} {Phys. Rev. B}\ }\textbf {\bibinfo {volume} {101}},\ \bibinfo
  {pages} {245303} (\bibinfo {year} {2020})}\BibitemShut {NoStop}%
\bibitem [{\citenamefont {Grigoriev}\ and\ \citenamefont
  {Ziman}(2017)}]{Grigoriev2017}%
  \BibitemOpen
  \bibfield  {author} {\bibinfo {author} {\bibfnamefont {P.~D.}\ \bibnamefont
  {Grigoriev}}\ and\ \bibinfo {author} {\bibfnamefont {T.}~\bibnamefont
  {Ziman}},\ }\bibfield  {title} {\bibinfo {title} {Magnetic oscillations
  measure interlayer coupling in cuprate superconductors},\ }\href
  {https://doi.org/10.1103/PhysRevB.96.165110} {\bibfield  {journal} {\bibinfo
  {journal} {Phys. Rev. B}\ }\textbf {\bibinfo {volume} {96}},\ \bibinfo
  {pages} {165110} (\bibinfo {year} {2017})}\BibitemShut {NoStop}%
\bibitem [{\citenamefont {Grigoriev}\ \emph {et~al.}(2016)\citenamefont
  {Grigoriev}, \citenamefont {Sinchenko}, \citenamefont {Lejay}, \citenamefont
  {Hadj-Azzem}, \citenamefont {Balay}, \citenamefont {Leynaud}, \citenamefont
  {Zverev},\ and\ \citenamefont {Monceau}}]{Grigoriev2016}%
  \BibitemOpen
  \bibfield  {author} {\bibinfo {author} {\bibfnamefont {P.~D.}\ \bibnamefont
  {Grigoriev}}, \bibinfo {author} {\bibfnamefont {A.~A.}\ \bibnamefont
  {Sinchenko}}, \bibinfo {author} {\bibfnamefont {P.}~\bibnamefont {Lejay}},
  \bibinfo {author} {\bibfnamefont {A.}~\bibnamefont {Hadj-Azzem}}, \bibinfo
  {author} {\bibfnamefont {J.}~\bibnamefont {Balay}}, \bibinfo {author}
  {\bibfnamefont {O.}~\bibnamefont {Leynaud}}, \bibinfo {author} {\bibfnamefont
  {V.~N.}\ \bibnamefont {Zverev}},\ and\ \bibinfo {author} {\bibfnamefont
  {P.}~\bibnamefont {Monceau}},\ }\bibfield  {title} {\bibinfo {title} {Bilayer
  splitting versus fermi-surface warping as an origin of slow oscillations of
  in-plane magnetoresistance in rare-earth tritellurides},\ }\href
  {https://doi.org/10.1140/epjb/e2016-70159-6} {\bibfield  {journal} {\bibinfo
  {journal} {The European Physical Journal B}\ }\textbf {\bibinfo {volume}
  {89}},\ \bibinfo {pages} {151} (\bibinfo {year} {2016})}\BibitemShut
  {NoStop}%
\bibitem [{\citenamefont {Kartsovnik}\ \emph {et~al.}(2002)\citenamefont
  {Kartsovnik}, \citenamefont {Grigoriev}, \citenamefont {Biberacher},
  \citenamefont {Kushch},\ and\ \citenamefont {Wyder}}]{Kartsovnik2002}%
  \BibitemOpen
  \bibfield  {author} {\bibinfo {author} {\bibfnamefont {M.~V.}\ \bibnamefont
  {Kartsovnik}}, \bibinfo {author} {\bibfnamefont {P.~D.}\ \bibnamefont
  {Grigoriev}}, \bibinfo {author} {\bibfnamefont {W.}~\bibnamefont
  {Biberacher}}, \bibinfo {author} {\bibfnamefont {N.~D.}\ \bibnamefont
  {Kushch}},\ and\ \bibinfo {author} {\bibfnamefont {P.}~\bibnamefont
  {Wyder}},\ }\bibfield  {title} {\bibinfo {title} {Slow oscillations of
  magnetoresistance in quasi-two-dimensional metals},\ }\href
  {https://doi.org/10.1103/PhysRevLett.89.126802} {\bibfield  {journal}
  {\bibinfo  {journal} {Phys. Rev. Lett.}\ }\textbf {\bibinfo {volume} {89}},\
  \bibinfo {pages} {126802} (\bibinfo {year} {2002})}\BibitemShut {NoStop}%
\bibitem [{\citenamefont {Bastin}\ \emph {et~al.}(1971)\citenamefont {Bastin},
  \citenamefont {Lewiner}, \citenamefont {Betbeder-matibet},\ and\
  \citenamefont {Nozieres}}]{Bastin1971}%
  \BibitemOpen
  \bibfield  {author} {\bibinfo {author} {\bibfnamefont {A.}~\bibnamefont
  {Bastin}}, \bibinfo {author} {\bibfnamefont {C.}~\bibnamefont {Lewiner}},
  \bibinfo {author} {\bibfnamefont {O.}~\bibnamefont {Betbeder-matibet}},\ and\
  \bibinfo {author} {\bibfnamefont {P.}~\bibnamefont {Nozieres}},\ }\bibfield
  {title} {\bibinfo {title} {Quantum oscillations of the hall effect of a
  fermion gas with random impurity scattering},\ }\href
  {https://doi.org/https://doi.org/10.1016/S0022-3697(71)80147-6} {\bibfield
  {journal} {\bibinfo  {journal} {Journal of Physics and Chemistry of Solids}\
  }\textbf {\bibinfo {volume} {32}},\ \bibinfo {pages} {1811} (\bibinfo {year}
  {1971})}\BibitemShut {NoStop}%
\bibitem [{\citenamefont {Castro~Neto}\ \emph {et~al.}(2009)\citenamefont
  {Castro~Neto}, \citenamefont {Guinea}, \citenamefont {Peres}, \citenamefont
  {Novoselov},\ and\ \citenamefont {Geim}}]{Castro2009}%
  \BibitemOpen
  \bibfield  {author} {\bibinfo {author} {\bibfnamefont {A.~H.}\ \bibnamefont
  {Castro~Neto}}, \bibinfo {author} {\bibfnamefont {F.}~\bibnamefont {Guinea}},
  \bibinfo {author} {\bibfnamefont {N.~M.~R.}\ \bibnamefont {Peres}}, \bibinfo
  {author} {\bibfnamefont {K.~S.}\ \bibnamefont {Novoselov}},\ and\ \bibinfo
  {author} {\bibfnamefont {A.~K.}\ \bibnamefont {Geim}},\ }\bibfield  {title}
  {\bibinfo {title} {The electronic properties of graphene},\ }\href
  {https://doi.org/10.1103/RevModPhys.81.109} {\bibfield  {journal} {\bibinfo
  {journal} {Rev. Mod. Phys.}\ }\textbf {\bibinfo {volume} {81}},\ \bibinfo
  {pages} {109} (\bibinfo {year} {2009})}\BibitemShut {NoStop}%
\bibitem [{\citenamefont {Gorbar}\ \emph {et~al.}(2002)\citenamefont {Gorbar},
  \citenamefont {Gusynin}, \citenamefont {Miransky},\ and\ \citenamefont
  {Shovkovy}}]{Gorbar2002}%
  \BibitemOpen
  \bibfield  {author} {\bibinfo {author} {\bibfnamefont {E.~V.}\ \bibnamefont
  {Gorbar}}, \bibinfo {author} {\bibfnamefont {V.~P.}\ \bibnamefont {Gusynin}},
  \bibinfo {author} {\bibfnamefont {V.~A.}\ \bibnamefont {Miransky}},\ and\
  \bibinfo {author} {\bibfnamefont {I.~A.}\ \bibnamefont {Shovkovy}},\
  }\bibfield  {title} {\bibinfo {title} {Magnetic field driven metal-insulator
  phase transition in planar systems},\ }\href
  {https://doi.org/10.1103/PhysRevB.66.045108} {\bibfield  {journal} {\bibinfo
  {journal} {Phys. Rev. B}\ }\textbf {\bibinfo {volume} {66}},\ \bibinfo
  {pages} {045108} (\bibinfo {year} {2002})}\BibitemShut {NoStop}%
\bibitem [{\citenamefont {Gusynin}\ and\ \citenamefont
  {Sharapov}(2005)}]{Gusynin2005}%
  \BibitemOpen
  \bibfield  {author} {\bibinfo {author} {\bibfnamefont {V.~P.}\ \bibnamefont
  {Gusynin}}\ and\ \bibinfo {author} {\bibfnamefont {S.~G.}\ \bibnamefont
  {Sharapov}},\ }\bibfield  {title} {\bibinfo {title} {Magnetic oscillations in
  planar systems with the dirac-like spectrum of quasiparticle excitations. ii.
  transport properties},\ }\href {https://doi.org/10.1103/PhysRevB.71.125124}
  {\bibfield  {journal} {\bibinfo  {journal} {Phys. Rev. B}\ }\textbf {\bibinfo
  {volume} {71}},\ \bibinfo {pages} {125124} (\bibinfo {year}
  {2005})}\BibitemShut {NoStop}%
\bibitem [{\citenamefont {K\"uppersbusch}\ and\ \citenamefont
  {Fritz}(2017)}]{Kueppersbusch2017}%
  \BibitemOpen
  \bibfield  {author} {\bibinfo {author} {\bibfnamefont {C.}~\bibnamefont
  {K\"uppersbusch}}\ and\ \bibinfo {author} {\bibfnamefont {L.}~\bibnamefont
  {Fritz}},\ }\bibfield  {title} {\bibinfo {title} {Modifications of the
  lifshitz-kosevich formula in two-dimensional dirac systems},\ }\href
  {https://doi.org/10.1103/PhysRevB.96.205410} {\bibfield  {journal} {\bibinfo
  {journal} {Phys. Rev. B}\ }\textbf {\bibinfo {volume} {96}},\ \bibinfo
  {pages} {205410} (\bibinfo {year} {2017})}\BibitemShut {NoStop}%
\bibitem [{\citenamefont {Shishido}\ \emph {et~al.}(2018)\citenamefont
  {Shishido}, \citenamefont {Yamada}, \citenamefont {Sugii}, \citenamefont
  {Shimozawa}, \citenamefont {Yanase},\ and\ \citenamefont
  {Yamashita}}]{Shishido2018}%
  \BibitemOpen
  \bibfield  {author} {\bibinfo {author} {\bibfnamefont {H.}~\bibnamefont
  {Shishido}}, \bibinfo {author} {\bibfnamefont {S.}~\bibnamefont {Yamada}},
  \bibinfo {author} {\bibfnamefont {K.}~\bibnamefont {Sugii}}, \bibinfo
  {author} {\bibfnamefont {M.}~\bibnamefont {Shimozawa}}, \bibinfo {author}
  {\bibfnamefont {Y.}~\bibnamefont {Yanase}},\ and\ \bibinfo {author}
  {\bibfnamefont {M.}~\bibnamefont {Yamashita}},\ }\bibfield  {title} {\bibinfo
  {title} {Anomalous change in the de haas--van alphen oscillations of
  {${\mathrm{CeCoIn}}_{5}$} at ultralow temperatures},\ }\href
  {https://doi.org/10.1103/PhysRevLett.120.177201} {\bibfield  {journal}
  {\bibinfo  {journal} {Phys. Rev. Lett.}\ }\textbf {\bibinfo {volume} {120}},\
  \bibinfo {pages} {177201} (\bibinfo {year} {2018})}\BibitemShut {NoStop}%
\bibitem [{\citenamefont {Klotz}\ \emph {et~al.}(2018)\citenamefont {Klotz},
  \citenamefont {G\"otze}, \citenamefont {Sheikin}, \citenamefont {F\"orster},
  \citenamefont {Graf}, \citenamefont {Park}, \citenamefont {Choi},
  \citenamefont {Hu}, \citenamefont {Petrovic}, \citenamefont {Wosnitza},\ and\
  \citenamefont {Green}}]{Klotz2018}%
  \BibitemOpen
  \bibfield  {author} {\bibinfo {author} {\bibfnamefont {J.}~\bibnamefont
  {Klotz}}, \bibinfo {author} {\bibfnamefont {K.}~\bibnamefont {G\"otze}},
  \bibinfo {author} {\bibfnamefont {I.}~\bibnamefont {Sheikin}}, \bibinfo
  {author} {\bibfnamefont {T.}~\bibnamefont {F\"orster}}, \bibinfo {author}
  {\bibfnamefont {D.}~\bibnamefont {Graf}}, \bibinfo {author} {\bibfnamefont
  {J.-H.}\ \bibnamefont {Park}}, \bibinfo {author} {\bibfnamefont {E.~S.}\
  \bibnamefont {Choi}}, \bibinfo {author} {\bibfnamefont {R.}~\bibnamefont
  {Hu}}, \bibinfo {author} {\bibfnamefont {C.}~\bibnamefont {Petrovic}},
  \bibinfo {author} {\bibfnamefont {J.}~\bibnamefont {Wosnitza}},\ and\
  \bibinfo {author} {\bibfnamefont {E.~L.}\ \bibnamefont {Green}},\ }\bibfield
  {title} {\bibinfo {title} {Fermi surface reconstruction and dimensional
  topology change in {Nd}-doped {CeCoIn$_{5}$}},\ }\href
  {https://doi.org/10.1103/PhysRevB.98.081105} {\bibfield  {journal} {\bibinfo
  {journal} {Phys. Rev. B}\ }\textbf {\bibinfo {volume} {98}},\ \bibinfo
  {pages} {081105} (\bibinfo {year} {2018})}\BibitemShut {NoStop}%
\bibitem [{\citenamefont {Dalgaard}\ \emph {et~al.}(2020)\citenamefont
  {Dalgaard}, \citenamefont {Lei}, \citenamefont {Wiedmann}, \citenamefont
  {Bremholm},\ and\ \citenamefont {Schoop}}]{Dalgaard2020}%
  \BibitemOpen
  \bibfield  {author} {\bibinfo {author} {\bibfnamefont {K.~J.}\ \bibnamefont
  {Dalgaard}}, \bibinfo {author} {\bibfnamefont {S.}~\bibnamefont {Lei}},
  \bibinfo {author} {\bibfnamefont {S.}~\bibnamefont {Wiedmann}}, \bibinfo
  {author} {\bibfnamefont {M.}~\bibnamefont {Bremholm}},\ and\ \bibinfo
  {author} {\bibfnamefont {L.~M.}\ \bibnamefont {Schoop}},\ }\bibfield  {title}
  {\bibinfo {title} {Anomalous shubnikov-de haas quantum oscillations in
  rare-earth tritelluride {${\mathrm{NdTe}}_{3}$}},\ }\href
  {https://doi.org/10.1103/PhysRevB.102.245109} {\bibfield  {journal} {\bibinfo
   {journal} {Phys. Rev. B}\ }\textbf {\bibinfo {volume} {102}},\ \bibinfo
  {pages} {245109} (\bibinfo {year} {2020})}\BibitemShut {NoStop}%
\bibitem [{\citenamefont {Xu}\ \emph {et~al.}(2020)\citenamefont {Xu},
  \citenamefont {Zhou}, \citenamefont {Wang}, \citenamefont {Wang},
  \citenamefont {Lin}, \citenamefont {Zeng}, \citenamefont {Cheng},
  \citenamefont {Weng},\ and\ \citenamefont {Xia}}]{ShengXu2020}%
  \BibitemOpen
  \bibfield  {author} {\bibinfo {author} {\bibfnamefont {S.}~\bibnamefont
  {Xu}}, \bibinfo {author} {\bibfnamefont {L.}~\bibnamefont {Zhou}}, \bibinfo
  {author} {\bibfnamefont {X.-Y.}\ \bibnamefont {Wang}}, \bibinfo {author}
  {\bibfnamefont {H.}~\bibnamefont {Wang}}, \bibinfo {author} {\bibfnamefont
  {J.-F.}\ \bibnamefont {Lin}}, \bibinfo {author} {\bibfnamefont {X.-Y.}\
  \bibnamefont {Zeng}}, \bibinfo {author} {\bibfnamefont {P.}~\bibnamefont
  {Cheng}}, \bibinfo {author} {\bibfnamefont {H.}~\bibnamefont {Weng}},\ and\
  \bibinfo {author} {\bibfnamefont {T.-L.}\ \bibnamefont {Xia}},\ }\bibfield
  {title} {\bibinfo {title} {Quantum oscillations and electronic structure in
  the large-chern-number topological chiral semimetal {PtGa}},\ }\href
  {https://doi.org/10.1088/0256-307X/37/10/107504} {\bibfield  {journal}
  {\bibinfo  {journal} {Chinese Physics Letters}\ }\textbf {\bibinfo {volume}
  {37}},\ \bibinfo {eid} {107504} (\bibinfo {year} {2020})}\BibitemShut
  {NoStop}%
\bibitem [{\citenamefont {Klemenz}\ \emph {et~al.}(2019)\citenamefont
  {Klemenz}, \citenamefont {Lei},\ and\ \citenamefont
  {Schoop}}]{klemenz2019topological}%
  \BibitemOpen
  \bibfield  {author} {\bibinfo {author} {\bibfnamefont {S.}~\bibnamefont
  {Klemenz}}, \bibinfo {author} {\bibfnamefont {S.}~\bibnamefont {Lei}},\ and\
  \bibinfo {author} {\bibfnamefont {L.~M.}\ \bibnamefont {Schoop}},\ }\bibfield
   {title} {\bibinfo {title} {Topological semimetals in square-net materials},\
  }\href@noop {} {\bibfield  {journal} {\bibinfo  {journal} {Annual Review of
  Materials Research}\ }\textbf {\bibinfo {volume} {49}},\ \bibinfo {pages}
  {185} (\bibinfo {year} {2019})}\BibitemShut {NoStop}%
\bibitem [{\citenamefont {McCann}\ and\ \citenamefont
  {Koshino}(2013)}]{mccann2013electronic}%
  \BibitemOpen
  \bibfield  {author} {\bibinfo {author} {\bibfnamefont {E.}~\bibnamefont
  {McCann}}\ and\ \bibinfo {author} {\bibfnamefont {M.}~\bibnamefont
  {Koshino}},\ }\bibfield  {title} {\bibinfo {title} {The electronic properties
  of bilayer graphene},\ }\href@noop {} {\bibfield  {journal} {\bibinfo
  {journal} {Reports on Progress in physics}\ }\textbf {\bibinfo {volume}
  {76}},\ \bibinfo {pages} {056503} (\bibinfo {year} {2013})}\BibitemShut
  {NoStop}%
\bibitem [{\citenamefont {McCann}\ and\ \citenamefont
  {Fal’ko}(2006)}]{mccann2006landau}%
  \BibitemOpen
  \bibfield  {author} {\bibinfo {author} {\bibfnamefont {E.}~\bibnamefont
  {McCann}}\ and\ \bibinfo {author} {\bibfnamefont {V.~I.}\ \bibnamefont
  {Fal’ko}},\ }\bibfield  {title} {\bibinfo {title} {Landau-level degeneracy
  and quantum hall effect in a graphite bilayer},\ }\href@noop {} {\bibfield
  {journal} {\bibinfo  {journal} {Physical review letters}\ }\textbf {\bibinfo
  {volume} {96}},\ \bibinfo {pages} {086805} (\bibinfo {year}
  {2006})}\BibitemShut {NoStop}%
\bibitem [{\citenamefont {Phinney}\ \emph {et~al.}(2021)\citenamefont
  {Phinney}, \citenamefont {Bandurin}, \citenamefont {Collignon}, \citenamefont
  {Dmitriev}, \citenamefont {Taniguchi}, \citenamefont {Watanabe},\ and\
  \citenamefont {Jarillo-Herrero}}]{Phinney2021_strong}%
  \BibitemOpen
  \bibfield  {author} {\bibinfo {author} {\bibfnamefont {I.~Y.}\ \bibnamefont
  {Phinney}}, \bibinfo {author} {\bibfnamefont {D.~A.}\ \bibnamefont
  {Bandurin}}, \bibinfo {author} {\bibfnamefont {C.}~\bibnamefont {Collignon}},
  \bibinfo {author} {\bibfnamefont {I.~A.}\ \bibnamefont {Dmitriev}}, \bibinfo
  {author} {\bibfnamefont {T.}~\bibnamefont {Taniguchi}}, \bibinfo {author}
  {\bibfnamefont {K.}~\bibnamefont {Watanabe}},\ and\ \bibinfo {author}
  {\bibfnamefont {P.}~\bibnamefont {Jarillo-Herrero}},\ }\bibfield  {title}
  {\bibinfo {title} {Strong interminivalley scattering in twisted bilayer
  graphene revealed by high-temperature magneto-oscillations},\ }\href
  {https://doi.org/10.1103/PhysRevLett.127.056802} {\bibfield  {journal}
  {\bibinfo  {journal} {Phys. Rev. Lett.}\ }\textbf {\bibinfo {volume} {127}},\
  \bibinfo {pages} {056802} (\bibinfo {year} {2021})}\BibitemShut {NoStop}%
\bibitem [{\citenamefont {Sunko}\ \emph {et~al.}(2017)\citenamefont {Sunko},
  \citenamefont {Rosner}, \citenamefont {Kushwaha}, \citenamefont {Khim},
  \citenamefont {Mazzola}, \citenamefont {Bawden}, \citenamefont {Clark},
  \citenamefont {Riley}, \citenamefont {Kasinathan}, \citenamefont {Haverkort}
  \emph {et~al.}}]{sunko2017maximal}%
  \BibitemOpen
  \bibfield  {author} {\bibinfo {author} {\bibfnamefont {V.}~\bibnamefont
  {Sunko}}, \bibinfo {author} {\bibfnamefont {H.}~\bibnamefont {Rosner}},
  \bibinfo {author} {\bibfnamefont {P.}~\bibnamefont {Kushwaha}}, \bibinfo
  {author} {\bibfnamefont {S.}~\bibnamefont {Khim}}, \bibinfo {author}
  {\bibfnamefont {F.}~\bibnamefont {Mazzola}}, \bibinfo {author} {\bibfnamefont
  {L.}~\bibnamefont {Bawden}}, \bibinfo {author} {\bibfnamefont
  {O.}~\bibnamefont {Clark}}, \bibinfo {author} {\bibfnamefont
  {J.}~\bibnamefont {Riley}}, \bibinfo {author} {\bibfnamefont
  {D.}~\bibnamefont {Kasinathan}}, \bibinfo {author} {\bibfnamefont
  {M.}~\bibnamefont {Haverkort}}, \emph {et~al.},\ }\bibfield  {title}
  {\bibinfo {title} {Maximal rashba-like spin splitting via
  kinetic-energy-coupled inversion-symmetry breaking},\ }\href@noop {}
  {\bibfield  {journal} {\bibinfo  {journal} {Nature}\ }\textbf {\bibinfo
  {volume} {549}},\ \bibinfo {pages} {492} (\bibinfo {year}
  {2017})}\BibitemShut {NoStop}%
\bibitem [{\citenamefont {Allocca}\ and\ \citenamefont
  {Cooper}(2021)}]{Allocca2021}%
  \BibitemOpen
  \bibfield  {author} {\bibinfo {author} {\bibfnamefont {A.~A.}\ \bibnamefont
  {Allocca}}\ and\ \bibinfo {author} {\bibfnamefont {N.~R.}\ \bibnamefont
  {Cooper}},\ }\href@noop {} {\bibinfo {title} {Low-frequency quantum
  oscillations from interactions in layered metals}} (\bibinfo {year} {2021}),\
  \Eprint {https://arxiv.org/abs/2103.08617} {arXiv:2103.08617
  [cond-mat.mes-hall]} \BibitemShut {NoStop}%
\bibitem [{\citenamefont {Allocca}\ and\ \citenamefont
  {Cooper}(2022)}]{allocca2022quantum}%
  \BibitemOpen
  \bibfield  {author} {\bibinfo {author} {\bibfnamefont {A.}~\bibnamefont
  {Allocca}}\ and\ \bibinfo {author} {\bibfnamefont {N.}~\bibnamefont
  {Cooper}},\ }\bibfield  {title} {\bibinfo {title} {Quantum oscillations in
  interaction-driven insulators},\ }\href@noop {} {\bibfield  {journal}
  {\bibinfo  {journal} {SciPost Physics}\ }\textbf {\bibinfo {volume} {12}},\
  \bibinfo {pages} {123} (\bibinfo {year} {2022})}\BibitemShut {NoStop}%
\bibitem [{\citenamefont {Allocca}\ and\ \citenamefont
  {Cooper}(2023)}]{allocca2023fluctuation}%
  \BibitemOpen
  \bibfield  {author} {\bibinfo {author} {\bibfnamefont {A.~A.}\ \bibnamefont
  {Allocca}}\ and\ \bibinfo {author} {\bibfnamefont {N.~R.}\ \bibnamefont
  {Cooper}},\ }\bibfield  {title} {\bibinfo {title} {Fluctuation-dominated
  quantum oscillations in excitonic insulators},\ }\href@noop {} {\bibfield
  {journal} {\bibinfo  {journal} {arXiv preprint arXiv:2302.06633}\ } (\bibinfo
  {year} {2023})}\BibitemShut {NoStop}%
\bibitem [{\citenamefont {Leeb}\ and\ \citenamefont {Knolle}()}]{code}%
  \BibitemOpen
  \bibfield  {author} {\bibinfo {author} {\bibfnamefont {V.}~\bibnamefont
  {Leeb}}\ and\ \bibinfo {author} {\bibfnamefont {J.}~\bibnamefont {Knolle}},\
  }\href {https://doi.org/10.5281/zenodo.8046745} {\bibinfo {title} {On the
  theory of difference frequency quantum oscillations}}\BibitemShut {NoStop}%
\bibitem [{\citenamefont {Alisultanov}\ \emph {et~al.}(2023)\citenamefont
  {Alisultanov}, \citenamefont {Abdullaev}, \citenamefont {Grigoriev},\ and\
  \citenamefont {Demirov}}]{Alisultanov2023}%
  \BibitemOpen
  \bibfield  {author} {\bibinfo {author} {\bibfnamefont {Z.~Z.}\ \bibnamefont
  {Alisultanov}}, \bibinfo {author} {\bibfnamefont {G.~O.}\ \bibnamefont
  {Abdullaev}}, \bibinfo {author} {\bibfnamefont {P.~D.}\ \bibnamefont
  {Grigoriev}},\ and\ \bibinfo {author} {\bibfnamefont {N.~A.}\ \bibnamefont
  {Demirov}},\ }\bibfield  {title} {\bibinfo {title} {Quantum oscillations of
  interlayer conductivity in a multilayer topological insulator},\ }\href
  {https://doi.org/10.1134/S106377612303010X} {\bibfield  {journal} {\bibinfo
  {journal} {Journal of Experimental and Theoretical Physics}\ }\textbf
  {\bibinfo {volume} {136}},\ \bibinfo {pages} {353} (\bibinfo {year}
  {2023})}\BibitemShut {NoStop}%
\end{thebibliography}%

\end{document}